\documentclass[fleqn,usenatbib]{mnras}
\usepackage{newtxtext,newtxmath}
\usepackage[T1]{fontenc}

\DeclareRobustCommand{\VAN}[3]{#2}
\let\VANthebibliography\thebibliography
\def\thebibliography{\DeclareRobustCommand{\VAN}[3]{##3}\VANthebibliography}

\outer\def\gtae {$\buildrel {\lower3pt\hbox{$>$}} \over 
{\lower2pt\hbox{$\sim$}} $}
\outer\def\ltae {$\buildrel {\lower3pt\hbox{$<$}} \over 
{\lower2pt\hbox{$\sim$}} $}

\newcommand{\tess}{\sl TESS}

\usepackage{graphicx}	
\usepackage{amsmath}	
\usepackage{url}

\title[Flares from low mass stars]{Searching for stellar flares from low mass stars using ASKAP and TESS}

\author[J. Rigney et al.]{Jeremy Rigney,$^{1,2,3}$\thanks{E-mail: jeremy.rigney@dias.ie}
Gavin Ramsay,$^{2}$
Eoin P. Carley,$^{1}$
J. Gerry Doyle,$^{2}$
Peter T. Gallagher,$^{1}$
\newauthor
Yuanming Wang,$^{4,5,6}$
Joshua Pritchard,$^{4,5,6}$
Tara Murphy,$^{4,6}$
Emil Lenc,$^{5}$
David L.~Kaplan$^{7}$
\\
$^{1}$Astronomy \& Astrophysics Section, DIAS Dunsink Observatory, Dublin Institute for Advanced Studies, Dublin, D15 XR2R, Ireland\\
$^{2}$Armagh Observatory and Planetarium, College Hill, Armagh, BT61 9DG, N. Ireland\\
$^{3}$School of Mathematics and Physics, Queen's University Belfast, University Road, Belfast, BT7 1NN, N. Ireland\\
$^{4}$Sydney Institute for Astronomy, School of Physics, University of Sydney, NSW 2006, Australia\\
$^{5}$Australia Telescope National Facility, CSIRO, Space and Astronomy, PO Box 76, Epping, NSW 1710, Australia\\
$^{6}$ARC Centre of Excellence for Gravitational Wave Discovery (OzGrav), Hawthorn, Victoria, Australia\\
$^{7}$Department of Physics, University of Wisconsin-Milwaukee, P.O. Box 413, Milwaukee, WI 53201, USA
}

\date{Accepted XXX. Received YYY; in original form ZZZ}

\pubyear{2022}

\begin{document}
\label{firstpage}
\pagerange{\pageref{firstpage}--\pageref{lastpage}}
\maketitle

\begin{abstract}
Solar radio emission at low frequencies ($<$1~GHz) can provide valuable information on processes driving flares and coronal mass ejections (CMEs). Radio emission has been detected from active M dwarf stars, suggestive of much higher levels of activity than previously thought. Observations of active M dwarfs at low frequencies can provide information on the emission mechanism for high energy flares and possible stellar CMEs. Here, we conducted two observations with the Australian Square Kilometre Pathfinder Telescope (ASKAP) totalling 26 hours and scheduled to overlap with the Transiting Exoplanet Survey Satellite (TESS) Sector 36 field, utilising the wide fields of view of both telescopes to search for multiple M dwarfs. We detected variable radio emission in Stokes I centered at 888\;MHz from four known active M dwarfs. Two of these sources were also detected with Stokes V circular polarisation. When examining the detected radio emission characteristics, we were not able distinguish between the models for either electron cyclotron maser or gyrosynchrotron emission. These detections add to the growing number of M dwarfs observed with variable low frequency emission.

\end{abstract}



\section{Introduction}

M dwarfs are core hydrogen-burning stars with masses between 0.07 -- 0.6 M$\sun$ and surface temperatures ranging from 2500\;K (M9.5) -- 3700\;K (M0) \citep{rajpurohit2013effective}. They are the most common type of star in the Milky Way, making up an estimated 70\% of the stellar population \citep{henry2006solarneighbourhood, bochanski2010luminosity}. High rotation rates (< 10 days) and convective cores, particularly in late type M dwarfs (> M4), generate complex powerful global magnetic fields with large stable star spots \citep{kochukhov2021, Gunther2020}. Hence these stars can regularly produce flares with energies in excess of $\sim10^{32}$ ergs up to $\sim10^{35}$ ergs and superflares with energies surpassing $10^{36}$ ergs \citep{Osten2010, JDavenport2016, Schmidt2019, ramsay2021}. It is thought these stellar flares occur via the same magnetic reconnection process as observed in solar flares, and this process also leads to the expulsion of plasma into the stellar atmosphere, known as a stellar coronal mass ejection (CME). While CMEs are a regular occurrence on the Sun, there are relatively few observational confirmations of stellar CMEs. A potential means of observing stellar flares and CMEs is through radio observations. Although there has been observations of stars which are consistent with being caused by CMEs, there has been no unambiguous detection of a Type II like burst or CME signature on a star other than the Sun (e.g. \citet{mullan2019, Zic2020}). Routine detections of such radio bursts would open up new avenues of flare and CME research on M-dwarf stars, and new-fields of exo-space weather.

On the Sun, flares and CMEs are associated with the acceleration of energetic electrons, producing emission from gamma-ray to radio. The mildly relativistic electrons can often be associated with bursts of radio emission. For example, Type III bursts are produced from mildly relativistic electrons escaping on open magnetic fields \citep{reid2014review}, Type IVs are associated with energetic electrons trapped in either the flare or CME \citep{carley2021observations}, while Type II bursts are generated from CME-driven shocks \citep{nelson1985type, maguire2020, morosan2021moving}. Given the fact these radio bursts can be coherent emission, they have flux densities averaging $10^8$ Jy and can be as high as $10^{11}$ Jy \citep{cerruti2008effect, raja2022spectral}. Radio bursts can be used as a diagnostic of plasma density, plasma turbulence, electron energies and the sites of particle acceleration during the eruptive process. In other cases the radio bursts can be associated with gyrosynchrotron emission (particularly the Type IVs; \citealt{carley2017estimation}) and in rarer cases the flares and CMEs have also been known to be associated with Electron Cyclotron Maser (ECM) emission \citep{morosan2019variable, carley2019loss, liu2018solar}. 

The close association of solar radio bursts with the flare-CME energy release process, and the high intensity of the coherent radio emission, has prompted many searches for analogous activity on Sun-like and M-dwarf stars. Searches for such radio activity have tended to concentrate efforts at higher frequencies e.g., 1\;GHz with telescopes such as the VLA or Arecibo \citep{stepanov2001microwave, osten2006wide, Villadsen2019}, with some success in detecting bursts with flux densities of 10\textemdash 100 mJy. Only a handful of studies have been able to detect such bursts at lower frequencies, e.g. the detection of 60 mJy radio bursts at 154 MHz from UV Ceti using MWA \citep{Lynch2017}, or 21 mJy radio bursts showing drifting features using VLA P-band (230-470 MHz) observations \citep{Crosley2018}. Further, \cite{Callingham2021_crdra} observed M dwarfs with peak radio fluxes reaching 205\;mJy at 170 MHz. Recently, data from telescopes such as the LOw Frequency ARray (LOFAR) (e.g. \cite{Callingham2021pop}) and the Australian Square Kilometre Pathfinder Telescope (ASKAP) (e.g. \cite{Zic2020}) have been used to search for stellar-analogues to Type II and IV emission from M dwarfs.

\begin{table*}
	\caption{The observation log of the ASKAP data which were used in this study.}
	\label{askaplog}
	\begin{tabular}{ccccccc} 
		\hline
		Obs. ID & Obs. Start (UT) & Obs. End (UT) & Duration (hr) & Freq. Range (MHz) & Freq. Res. (MHz) & Time Res. (sec)\\
		\hline
		SB25035 & 2021-03-20 03:10:58 & 2021-03-20 16:00:31 & 12.826 & 700 -1000 & 1 & 10 \\
		SB25077 & 2021-03-21 03:36:49 & 2021-03-21 16:26:22 & 12.826 & 700 -1000 & 1 & 10 \\
		\hline
	\end{tabular}
\end{table*}

In terms of the expected emission mechanism of stellar radio bursts, much of the current research observing M dwarfs at low frequencies has centred on ECM. This beamed coherent emission can be characterised by a high degree of circular polarisation with a high (> $10^{12}$ K) brightness temperature \citep{kellermann1969spectra, Melrose1982,dulk1985radio, vedantham2021mechanism}. This is thought to emerge from flaring activity \citep{Lynch2017, Zic2020} or auroral processes \citep{Kuz2012, Met2017, Callingham2021_crdra, Zic2019} and its characteristics help to better understand the radio emission sources and diagnostics such as magnetic fields of M dwarfs \citep{Callingham2021pop}. Recent detections of circularly polarised long duration radio emission from a quiescent M dwarf may even be indicative of star-planet interactions, similar to the Jupiter-Io system \citep{vedantham2020coherent} (see \citealt{marques2017} for a detailed overview of Jupiter decametric emission). Recently more comprehensive studies utilising survey data have yielded statistics from an increasingly large number of M dwarfs, further helping to constrain and categorise the dominant emission mechanisms of these active stars. \cite{Villadsen2019} detected 22 radio bursts from 5 active M Dwarfs using the VLA, concluding that all bursts were likely produced via a coherent emission process due to high (>50\%) circular polarisation. 

\cite{Callingham2021pop} detected 19 M dwarfs in the LOFAR Two Meter Sky Survey (LoTSS), all with high degrees of circular polarisation and high brightness temperatures indicative of either plasma emission or ECM. They highlight that gigahertz detections typically fall into two categories, either long-duration low polarisation or polarised short-duration bursts. It is clear that M dwarfs can produce low frequency radio emission through high energy events related to flares and active regions on the stellar surface. 

\cite{pritchard2021} conducted a search for circularly polarised stars within the Rapid ASKAP Continuum Survey (RACS, \citep{mcconnell2020RACS}), detecting among their results 9 M dwarfs with emission centered at 887.5 MHz and a 288 MHz bandwidth. The majority of these sources had high (>50\%) circular polarisation. Combined with high brightness temperature calculations, the authors concluded that coherent processes likely drive the emission from these sources.

Active M dwarf CMEs are expected to occur much more frequently when compared to the solar flare-CME scale \citep{Crosley2018, moschou2019stellar}. Hence the relatively low number of detections of stellar radio bursts is surprising \citep{crosley2016search}.
It is speculated that the high magnetic field strength of M dwarfs are high enough to constrain plasma release \citep{alvarado2018, kochukhov2021}, or that the local Alfvén speeds are too fast for CME shocks to occur \citep{Villadsen2019}. This would inhibit the production of shock-related radio emission, for example. Some stellar CME signatures may still be detectable at radio frequencies, albeit as a faint long duration gradual brightening rather than the more recognised emission of a solar type II radio burst \citep{alvarado2019coronal}.
Further exploration of this theory suggests the radio emission may be pushed beyond detectable limits with current ground based radio telescopes. The magnetic suppression could push the shock into the lower density outer corona and the resulting Type II-analogous emission is <20 MHz and below 1 mJy (below the ionospheric cutoff and sensitivity limits of current radio telescopes) and likely not-detectable \citep{alvarado2020tuning}. Recent simulations hint at detectable radio emission from high energy stellar CMEs at very low frequencies \citep{fionnagain2022coronal}, yet sustaining regular plasma ejections yields high mass loss rates \citep{osten2015connecting}. This conflicts with the long projected lifetimes of low mass stars.

Multi-wavelength observations of M dwarf activity has helped paint a clearer picture of flare energies, emission mechanisms, magnetic field strengths \citep{murphy2013vast}, star spot coverage, stellar activity-rotation period relation as well as many other factors which can help in understanding space weather in possible exoplanetary systems. 


With more detections of M dwarfs in the low frequency radio regime, a number of possible emission mechanisms have been suggested. 
\cite{pope2021tess} examined TESS light curves of the 19 M dwarfs detected with LOFAR \cite{Callingham2021pop} and found that for M dwarfs showing no flare activity but significant radio emission, magnetic processes were more likely to be driving the radio emission rather than coronal processes. Two targets were contemporaneously observed with LOFAR and TESS for 8 hours, although no flares were detected in either the optical or radio bands during this time.

However, the relation between radio emission and activity on these stars requires more investigation, particularly examining the connection between energy release across the electromagnetic spectrum. The aim of this work was to utilise simultaneous ASKAP and Transiting Exoplanet Survey Satellite (TESS) observations of two regions in the Southern Hemisphere in the search for variable emission from multiple active M dwarfs. Observing at both radio and optical wavelengths allowed for any detected radio emission to be correlated with optical flaring activity or period of quiescence on the stars. We made use of TESS Sector 36 observations which contain pointings at latitudes suitable for simultaneous observations with the ASKAP radio telescope. This work builds on previous observations of M dwarf observations at <1 GHz, such as \citet{pritchard2021}, \citet{Zic2019}, \citet{Zic2020}, and \citet{Villadsen2019}.

In Section 2 the observations and data reduction is described for both data sets. In Section 3 we report on the results of the observations and detection of M dwarfs. Section 4 covers the discussion on our findings. Section 5 contains the summary and conclusions of the paper.

\section{Observations and Data Reduction}

\subsection{Australian Square Kilometre Array Pathfinder Telescope}

The Australian Square Kilometre Array Pathfinder (ASKAP, \citep{johnston2008, hotan2021australian}) telescope is a 36 dish radio interferometric array constructed in a radio quiet zone in Western Australia. The system observes from 700 -- 1800\,MHz. The array has a 6 kilometre maximum baseline, and has a 30 square degree field of view. ASKAP is primarily designated as a survey telescope, with a current survey mapping radio sources in the southern sky to a sensitivity of < 0.25\;mJy \citep{mcconnell2020RACS}. For the current study, Directors Discretionary Time was used to take advantage of the overlapping observing region with TESS (see \S \ref{sec:TESS}).

Observations were scheduled for the dates 20-03-2021 (ID SB25035) and 21-03-2021 (ID SB 25077). The frequency range was centered at 888 MHz with a bandwidth of 288\,MHz, in 1\,MHz channels (see Table \ref{askaplog}). Each observation totalled 13 hours at 10-second time resolution with the data being initially processed using the current ASKAPsoft pipeline (see \citealt{Murphy2021, cornwell2012wide} for details). Further processing to produce images and dynamic spectra was completed using the \textit{Common Astronomy Software Applications} (\textit{CASA} package, \citealt{mcmullin2007casa}). Full 13-hour time-range integrations were used to produce Stokes $I$ and $V$ images with a resulting mean RMS$_{I} = 0.024$\;mJy beam$^{-1}$ and RMS$_{V} = 0.021$\;mJy beam$^{-1}$. The resulting 13 hour Stokes I images were then searched for sources using the \textit{Selavy} \citep{whiting2012source} routine.

\begin{figure*}
	\includegraphics[width=0.815\textwidth]{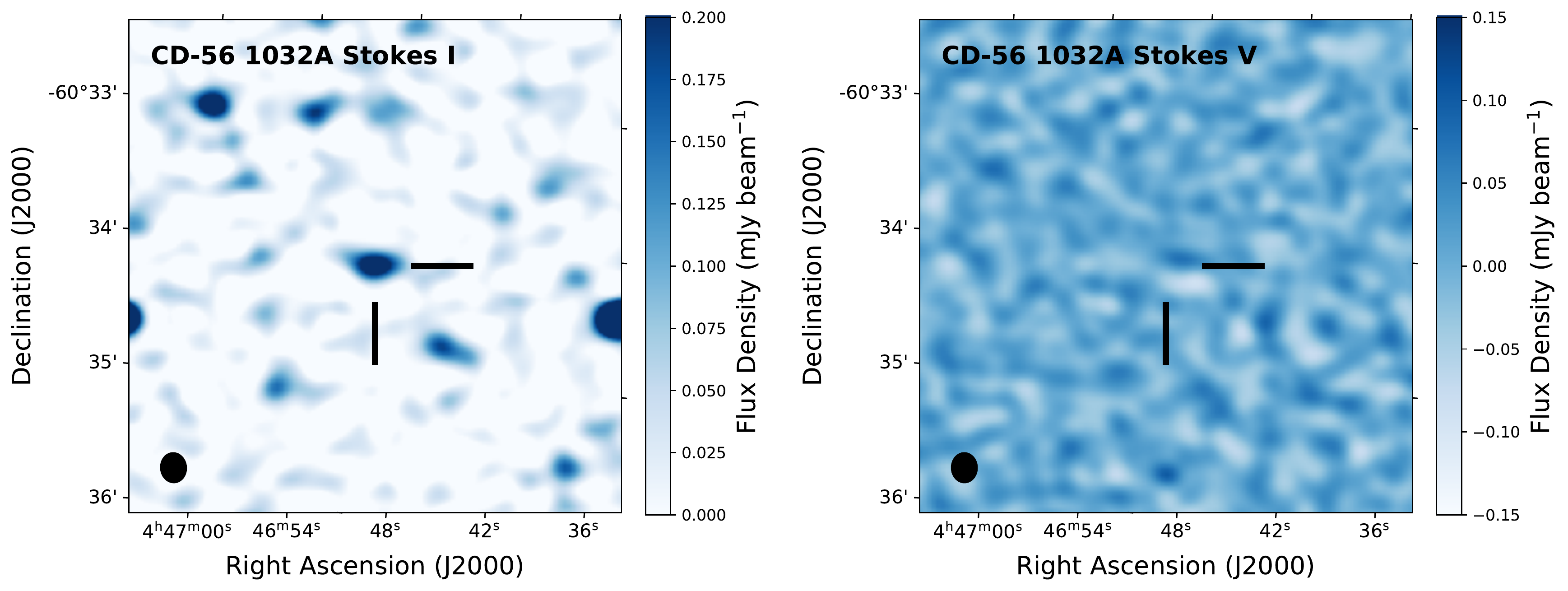}\\
	\includegraphics[width=0.815\textwidth]{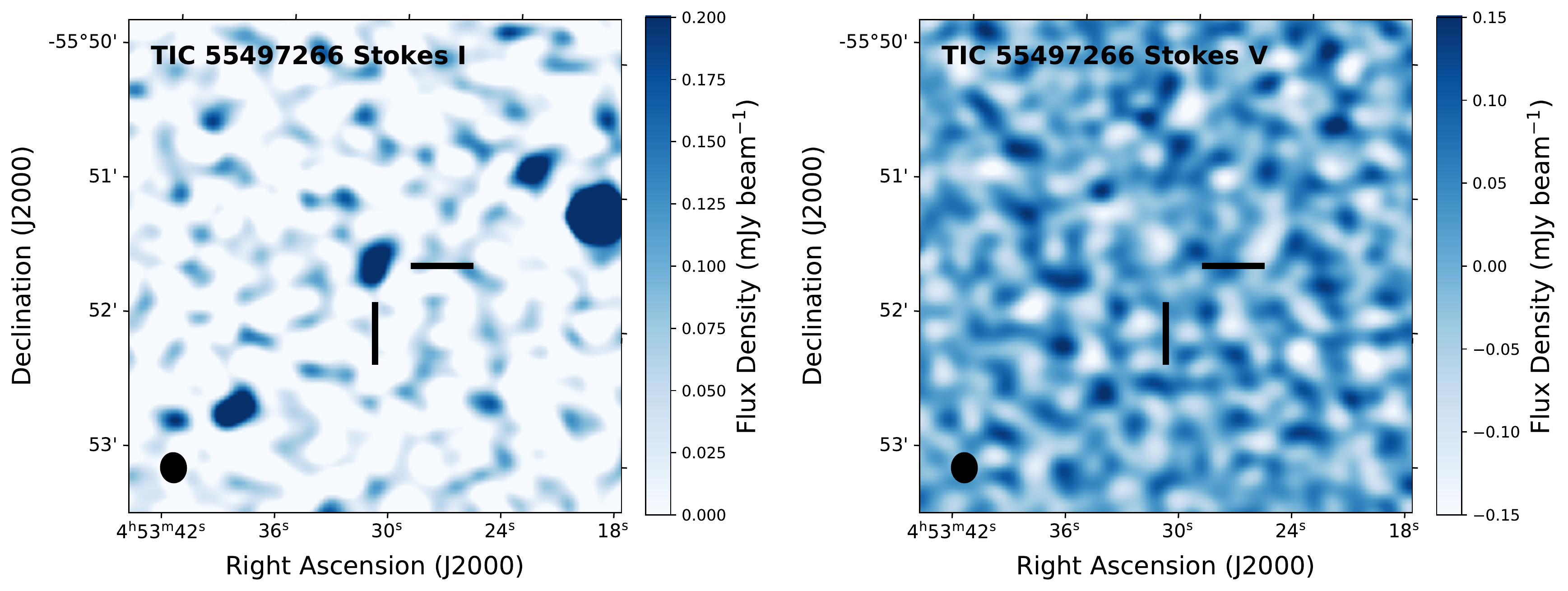}\\
	\includegraphics[width=0.815\textwidth]{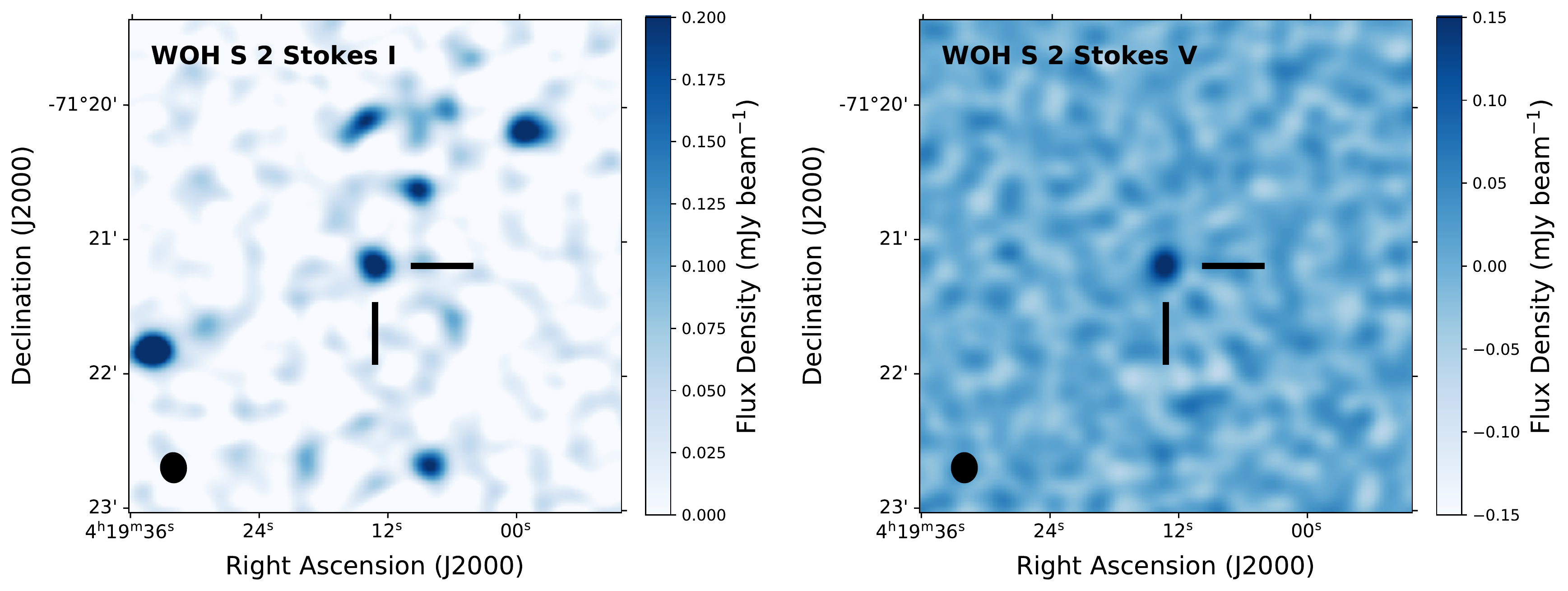}\\
	\includegraphics[width=0.815\textwidth]{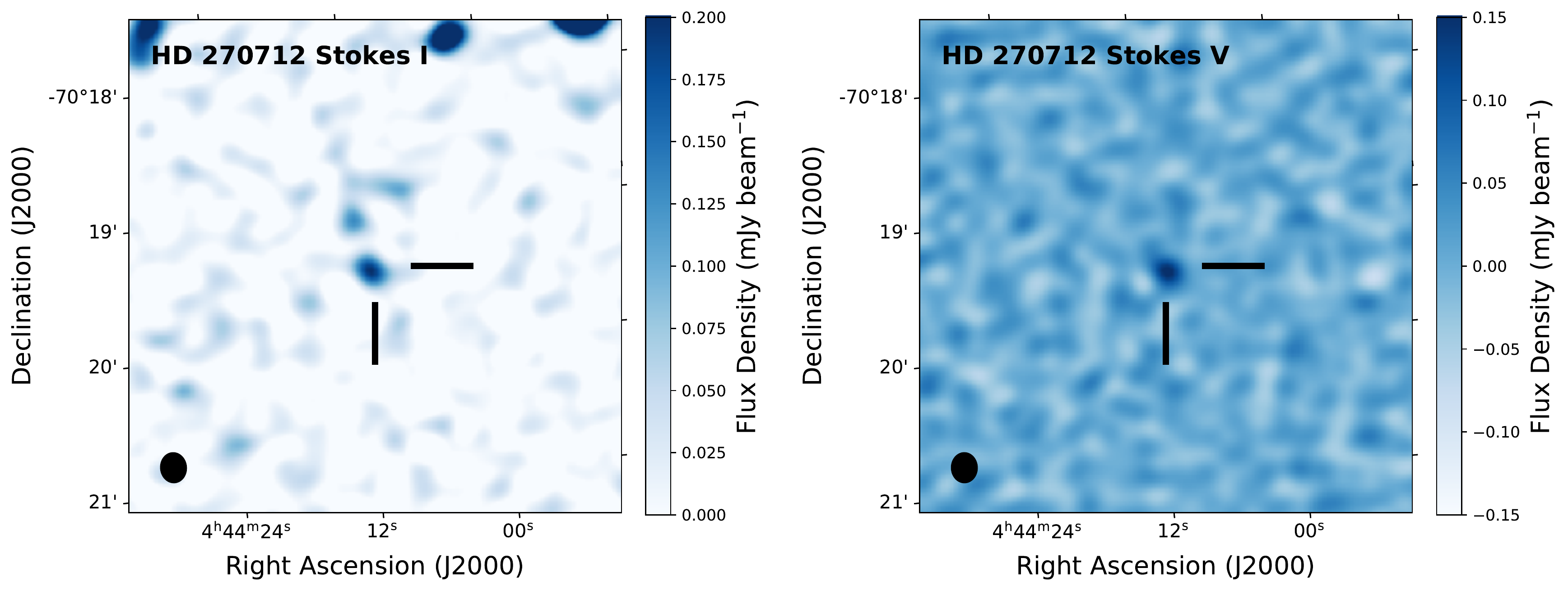}\\
    \caption{ASKAP 13 hour Stokes $I$ (left column) and $V$ (right column) images of the four detected source coordinates. Source locations are highlighted by cross-hairs, centered on \citet{Gunther2020} catalog coordinates (J2020, proper motion corrected). CD-56 1032A and TIC 55497266 were detected in Stokes $I$ only. WOH S 2 and HD 270212 were detected in both Stokes $I$ and positive Stokes $V$. The synthesised beam size is illustrated by the black ellipse in the bottom left of each image and is $13.9" \times 11.4"$.} 
    \label{fig:ASKAP_allstars}
\end{figure*}

The coordinates of significant ($>4\sigma$) sources were compared to the active M dwarf catalogue of \citet{Gunther2020}: four sources were identified from this processing (see Figure \ref{fig:ASKAP_allstars}. The calibrated beam visibilities containing the detected sources were then selected for further processing. Following the method of Section 2.2.1 in \citet{wang2021askap} a multi-frequency and multi-scale synthesis model image was created for each beam using the CASA \texttt{tclean} function. The model images were converted to model visibilities, then subtracted from the original calibrated visibilities. We imaged these model-subtracted visibilities across the full frequency range in 10 minute integrations which identified variable sources in the field and subtracted non-varying sources. The same process was repeated for Stokes V, Q and U. The polarisation leakage values were estimated to be approximately the same as that found in \citet{Zic2019} (i.e. $I$ to $V$: $\sim0.1\%$, $I$ to $Q$: $0.5\%$, $I$ to $U$: $1.5\%$). With sources in the center of the field, the effect of polarisation leakage is less that the effect from RMS noise. None of the sources detected were near the edge of either SB25035 or SB25077, thus the leakage is taken to be negligible when compared to the contribution of noise.

\begin{figure*}
	\includegraphics[width=0.33\textwidth]{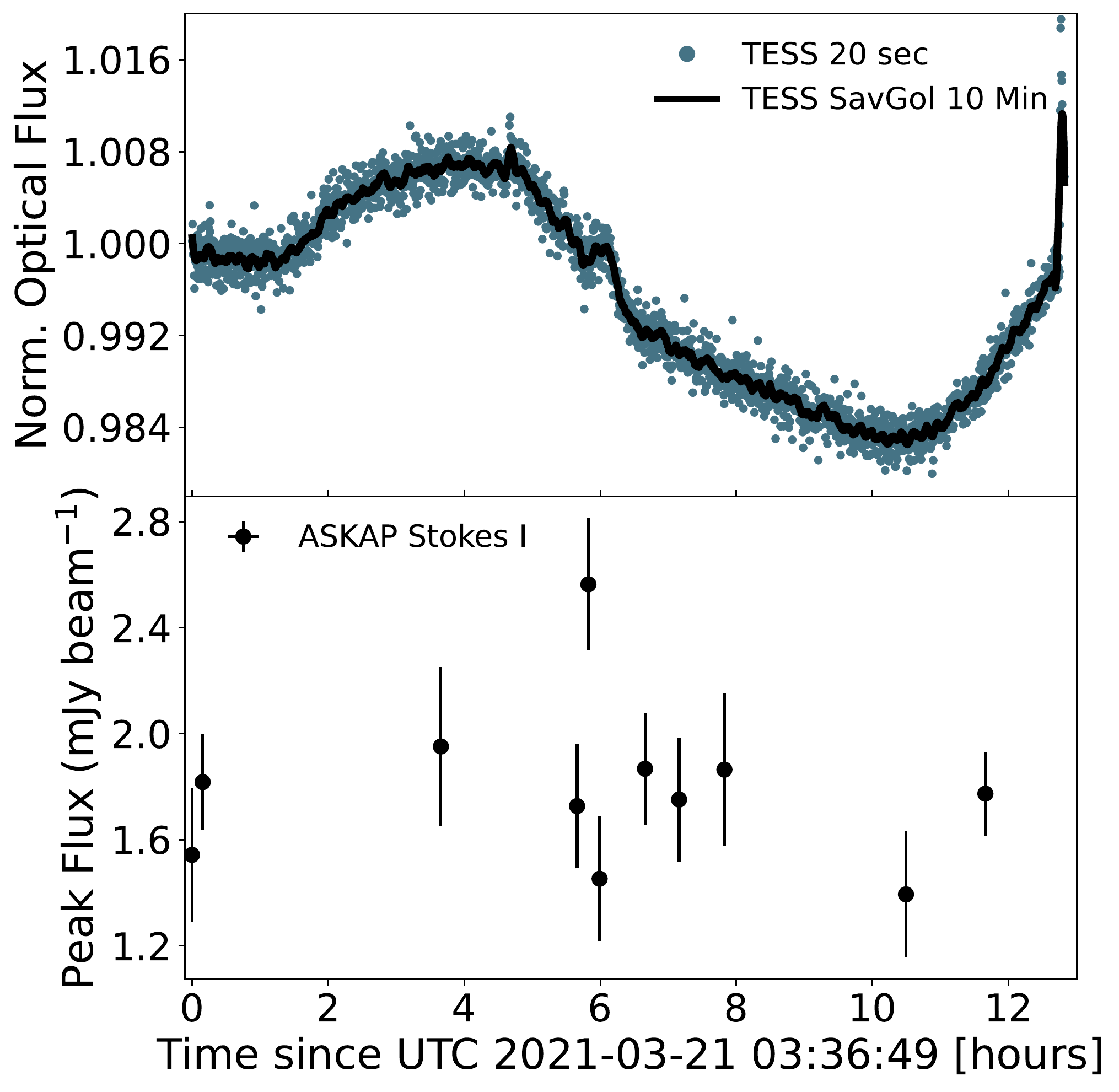}\hfill
	\includegraphics[width=0.33\textwidth]{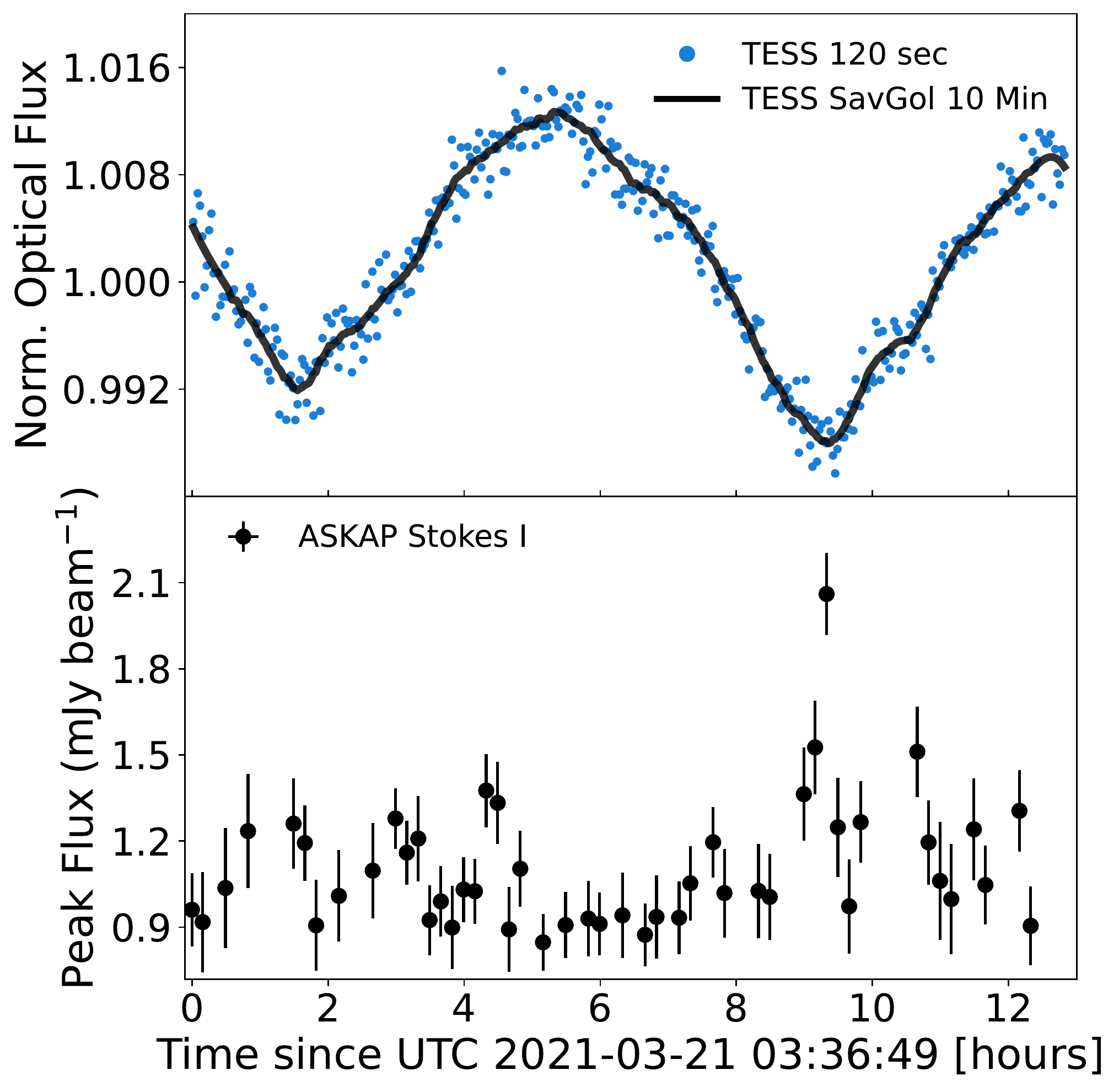}\hfill
	\includegraphics[width=0.33\textwidth]{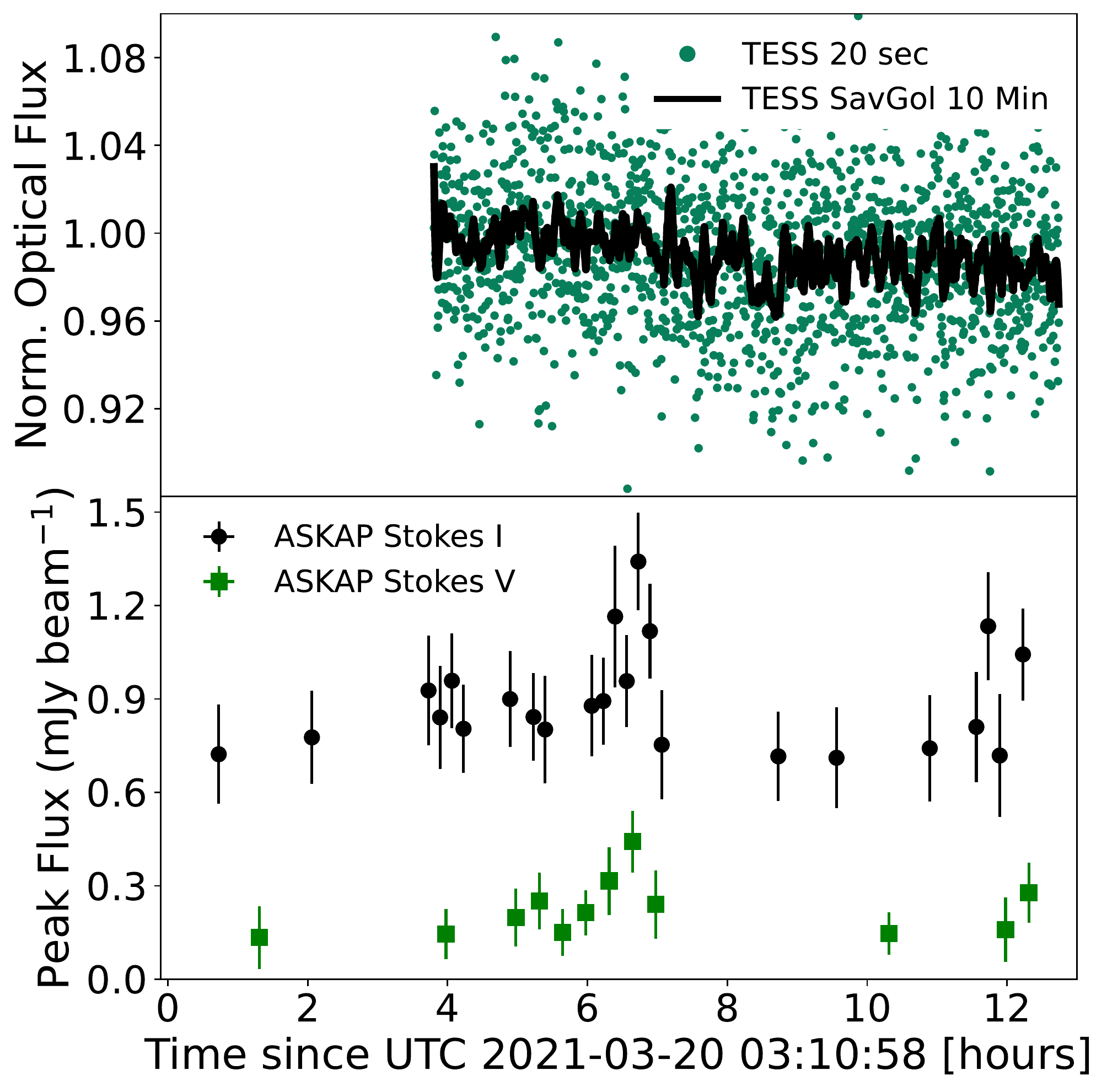}\hfill
    \caption{Simultaneous {\sl TESS} (upper panels) and {\sl ASKAP} (lower panels) Stokes $I$ data for CD-56 1032A (left panels); TIC 5549266 (middle panels) and WOH S 2 (right panels). The Stokes $I$ data is 4$\sigma$ above the RMS for all sources, and the Stokes $V$ for WOH S 2 is 2$\sigma$ above the RMS. The TESS (upper panel) y-axis is the normalised flux for each star.}
    \label{fig:tess_askap_all}
\end{figure*}

The peak calibrated flux density was obtained from the central pixel of the 2-D Gaussian at the source coordinates. This, with the addition of the model subtracted flux density, was used as the total flux density for an unresolved point source for each 10 minute image. The RMS for each image was also calculated and averaged across all images to determine the noise threshold for a detection. 
The results of this analysis is presented in Section \ref{results} (see Figure \ref{fig:tess_askap_all}). 

Dynamic spectra were created from visibility data using \textit{CASA}. The visibilities at each target coordinate were masked and a model of the field was created. This was then subtracted from the raw field visibilities, leaving only the target of interest visibilities. Phase center rotation, baseline averaging and time and frequency averaging were performed to generate the final dynamic spectra for the targets. The data were averaged down to 20 MHz frequency bands and 10 minute integrations. A circularly polarised radio burst was detected in the Stokes V dynamic spectrum for one of the stars.

\begin{figure*}
	\includegraphics[width=1\textwidth]{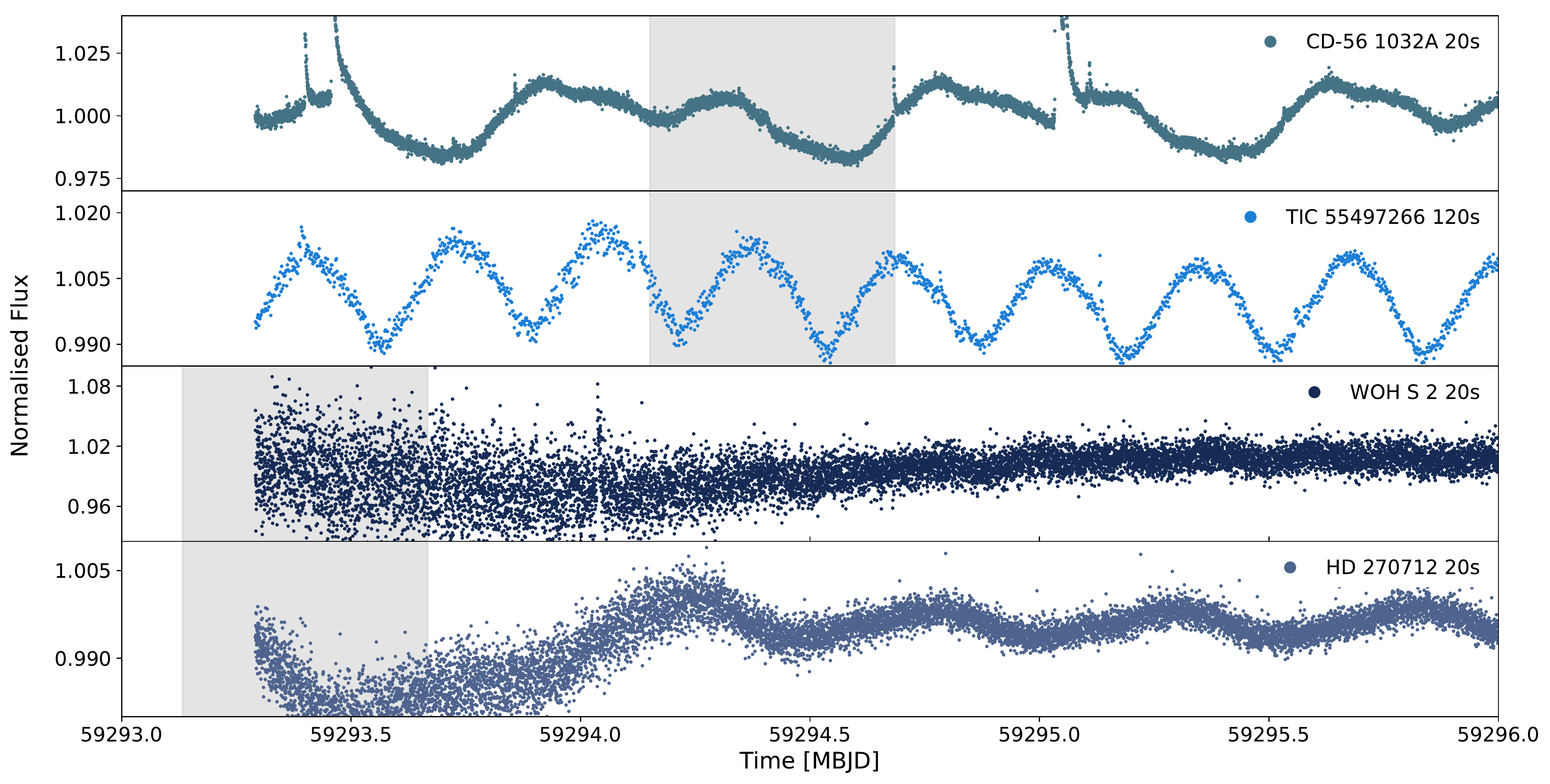}
    \caption{TESS Simple Aperture Photometry (SAP) optical light curves for the four detected M dwarfs. This figure illustrates the different characteristics of the light curves of each star, with clear periodic dips consistent with the presence of large star spots. The shaded region indicates when the stars were observed simultaneously with ASKAP and TESS. Note the missing WOH S 2 and HD 270712 data was due to a scheduled TESS data down-link between orbits 79 and 80 of Sector 36 observations. CD-56 1032A and TIC 55497266 were observed during ASKAP observing I.D. SB25077. WOH S 2 and HD 270712 were observed during I.D. SB25035. Only 120 second cadence TESS data was available for TIC 55497266. The region of HD 270712 co-observed with ASKAP has a very poor light curve severely impacted by instrumental effects, and was not used in this research. The light curve of WOH S 2 was similarly affected.}
    \label{fig:tess_LC_compare}
\end{figure*}

\begin{table*}
	\centering
	\caption{Stellar parameters of the four M dwarfs in our sample. Spectral types come from {\tt SIMBAD} apart from WOH S 2 which is the implied type from the Gaia EDR3 $(BP-RP)$ colour \citep{Gaia2021} and the relationships of \citet{PecautMamajek2013}. Temperature, radius, luminosity and mass are taken from the TESS Input Catalog \citep{Stassun2019} apart from WOH S 2 which come from the Gaia colour and mass-radius relationships from \citet{Baraffe2015} and \citet{Stassun2019,Gaia2021} data on low mass stars. The rotation periods and their error have been derived from all the available {\tess} data presented here.}
	\label{tab:star_info}
	\begin{tabular}{lcccc}
		\hline
		Star & CD-56 1032 A & TIC 55497266 & WOH S 2 & HD 270712 A/B\\
		\hline
		Other Name & TIC 220433363 & BPS CS 29520-0077 & TIC 29779873 & TIC 294750180 \\
		Classification & M3Ve & M1.5V & M4 & M0V/M1Ve \\
		Distance (pc) & 11.088$\pm$0.002 & 24.543$\pm$0.008 & 51.88$\pm$0.06 & 21.02$\pm$0.02 \\
		Effective Temp. (K) & 3400$\pm$160 & 3600$\pm$160 & 4090$\pm$200 & 3600$\pm$160 \\
		Radius ($R_{\odot}$) & 0.43$\pm$0.01 & 0.48$\pm$0.01 & 0.24$\pm$0.01 & - \\ 
		Luminosity ($L_{\odot}$) & 0.022 $\pm$ 0.005 & 0.034 $\pm$ 0.008 & 0.015$\pm$0.003 & - \\ 
		Mass ($M_{\odot}$) & 0.42$\pm$0.02 & 0.48$\pm$0.02 & 0.24$\pm$0.02 & - \\
		T mag & 8.7 & 10.0 & 12.5 & - \\
		Rotation Period (d) & 0.854$\pm$0.002 & 0.32442$\pm$0.00006 & 2.623$\pm$0.006 & 0.5254$\pm$0.0007 \\
		ASKAP Coords. ([RA] deg., [DEC] deg.) & (73.3802, -55.8597) & (71.7072, -60.5692) & (64.8055, -71.3532) & (71.0441, -70.3237)\\
		\hline
	\end{tabular}
\end{table*}

\begin{figure}
	\includegraphics[width=0.45\textwidth]{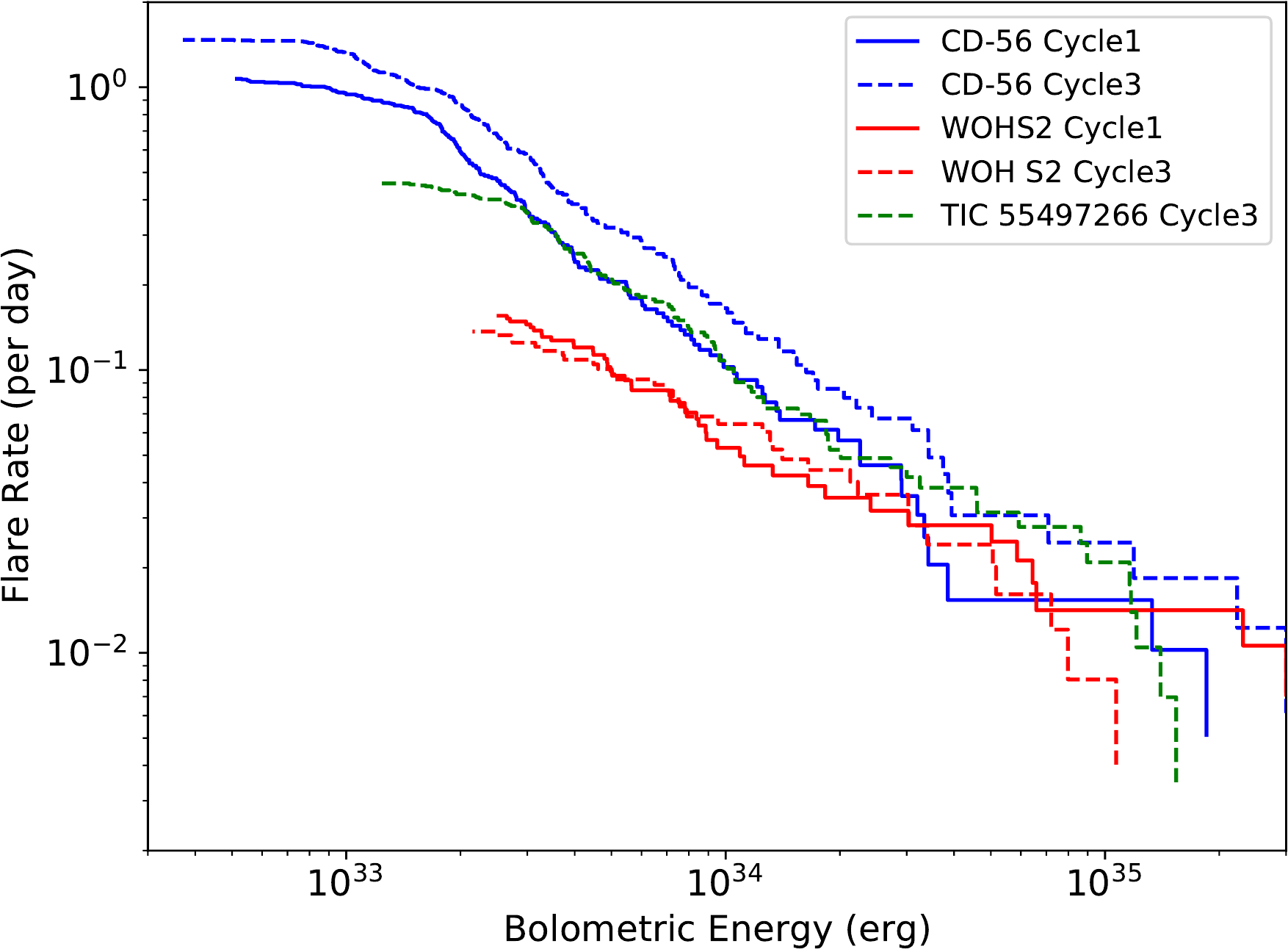}\hfill
    \caption{The flare frequency rate for the three stars in this study. It indicates how often a flare of a certain energy is observed. Two stars were observed in Cycles 1 and 3. These flare rates have been calculated from the first two months of TESS data.}
    \label{fig:tess_rate}
\end{figure}

\subsection{Transiting Exoplanet Survey Satellite (TESS)}
\label{sec:TESS} 

The Transiting Exoplanet Survey Satellite ({\tess}; \citealt{Ricker2015}) was launched in April 2018 with a primary mission of searching for exoplanets via the transit method around low mass stars. In its first two years, {\tess} completed a near all-sky survey observing more than 200,000 stars with a cadence of 2 minutes. In Year 3 a sample of stars were also observed with a cadence of 20 seconds. {\tess} observes in Sectors corresponding to observing one field for 26 days during two Earth orbits. Observations of Sector 36 took place between 7th March and 1st April 2021. The ASKAP observations were scheduled to align with Sector 36.

We downloaded the calibrated light curves for each of our four target stars from the MAST data archive\footnote{\url{https://archive.stsci.edu/tess/}}. We examined the light curve derived using the simple aperture photometry ({\tt SAP\_FLUX}) and the light curve which has been detrended and corrected for instrumental effects ({\tt PDCSAP\_FLUX}). Detrended points which did not have {\tt QUALITY=0} were removed and each light-curve was normalised by dividing the flux of each point by the mean flux of the star.

We show the {\tess} light curve of each source in Figure \ref{fig:tess_LC_compare} indicating the times where ASKAP data were obtained. Each light curve shows a clear modulation in flux over time which is very likely a signature of the stellar rotation period caused by starspots rotating into and out of view. For source WOH S 2 (TIC 29779873) and HD 270712 the observations at the start of the run were of relatively poor quality. 
We determined the period of the modulation using the Lomb Scargle periodogram and the Analysis of Variance deriving (consistent) periods of 0.8544 days (CD-56 1032A); 0.3244 days (TIC 55497266) and 2.62 days and 0.200 days in WOH S 2 (TIC 29779873).
Gaia shows there are two sources near WOH S 2 (G=14.2 and G=14.9) which are $1.28\arcsec$ apart and therefore fall into the same TESS pixel ($21\arcsec$ per pixel). One source has a Gaia (BP-RP) colour consistent with an M dwarf but the other star has no colour information. The {\tess} light curve of WOH S 2 is therefore a combination of both stars.
The rotation rate calculated from the TESS data for HD 270712 was 0.5264 days, however this cannot be attributed to either M dwarf in the visual binary.

There were no clear signatures of flares in the {\tess} light curves within the time-range of the ASKAP observations. However, there is one raised 'notch' like feature seen in CD-56 1032A (Figure \ref{fig:tess_LC_compare}, although it does not have a profile typical of flares). Comparing the light curve at the same rotational phase during the rest of the {\tess} observations (Figure \ref{fig:CD56_notch}), we note that flare events have been observed at this rotational phase (and others) during earlier and later rotations. 
We speculate that this notch like feature maybe related to active regions which are visible at this rotation phase, although it is unclear why other flare active regions do no show this feature.

Although there are no clear flares in the time interval where simultaneous ASKAP data was obtained we searched all the available {\tess} data for flares. 
To search for flares in the light curves, we removed the signature of the rotational modulation using a routine in the {\tt lightkurve} package \citep{lightkurve2018}. We then searched these flattened light curves for flares using the {\tt Altaipony}\footnote{\url{https://altaipony.readthedocs.io/en/latest}} suite of software which is an update of the {\tt Appaloosa}
\citep{JDavenport2016} software package. We obtained the stars bolometric luminosity from the TIC V8.0 catalogue \citep{Stassun2019} which uses Gaia DR2 \citep{gaia2018} and other information to determine the luminosities. (For WOH S 2 which does not have a luminosity in TIC V8 we used the Gaia (BP-RP) colour to estimate its luminosity by comparison with other M dwarfs of a similar colour). To determine the luminosity of the flares, we take the equivalent duration of the flare, calculated using {\tt Altaipony}, and multiply this by the stellar quiescent luminosity. We assume that the temperature of the flare is $\sim$12000 K and that the fraction of the emitted flux which falls within the {\tess} pass-band is $\sim$0.14 \citep{Schmitt2019}, thus requiring a multiplication of 7.1 to obtain the bolometric energy.

Using Cycle 1 and 3 {\tess} data, we determined the flare rate and energies and obtained the flare frequency distribution for three of the stars in each year of observation and this is shown in Figure \ref{fig:tess_rate}. There is good general agreement between the flare rate for the stars observed one year apart, with flares of energy 10$^{34}$ erg observed approximately every day in CD-56 1032A (unfortunately not during our ASKAP observations) and every tens of days of WOH S 2, with TIC 55497266 showing small flares every few days. 
There is a small decline in the number of flares seen in CD-56 1032A during Cycle 3 compared to Cycle 1. Since the stars have a range in brightness and distance, they are not equally sensitive to flares $<10^{34}$ erg. However, the rate of flares with energies $>$10$^{34}$ erg place CD-56 1032A and TIC 55497266 in the region where high energy flares may deplete the Ozone layer of any Earth-like planet in the host stars habitable zone (see for example \citealt{Gunther2020,ramsay2021}).

\section{Results}
\label{results}

\begin{table}
	\centering
	\caption{The minimum, maximum and mean detected radio flux density above 4 $\sigma$ from 10 minute integrations for each detected star. The table shows Stokes $I$ with Stokes $V$ in brackets for WOH S 2 and HD 270712. All values are given in mJy beam$^{-1}$.}
	\label{tab:fluxinfo}
	\begin{tabular}{lccc}
		\hline
		Star & Min. Flux & Max. Flux & Mean Flux $\pm 1\sigma$ \\
		\hline
		CD-56  & 1.38 & 2.89 & 1.72 $\pm$ 0.23 \\
		TIC 55 & 0.85 & 2.06 & 1.11 $\pm$ 0.14 \\
		WOH S 2 & 0.61 (0.24) & 1.34 (0.44) & 0.81 $\pm$ 0.07 (0.30 $\pm$ 0.07 LCP) \\
		HD 27 & 0 (0.37) & 0 (1.1) & 0 (0.54 $\pm$ 0.19 LCP)\\
		\hline
	\end{tabular}
\end{table}

There were 46 possible M dwarf targets identified from the \cite{Gunther2020} catalogue within the two regions observed. The comparison between radio source coordinates and Gaia coordinates (proper motion corrected) revealed four targets detected by ASKAP with TESS light curves.

Of these four sources detected in this study, two were detected in Stokes $I$ images only: CD-56 1032A and TIC 55497266. One source was detected in both Stokes $I$ and Stokes $V$: WOH S 2. One source was detected with faint quiescent Stokes I emission and a 100\% circularly polarised burst (HD 270712). There were no optical flares which coincided in time with radio events which were significantly above the mean Stokes $I$ or $V$ flux density. However, in CD-56 1032A, a radio peak coincided with a raised 'notch' like feature in the optical light curve: see Figure \ref{fig:tess_askap_all} for the ASKAP and {\tess} light curves. No discernible radio burst features were observed in the generated dynamic spectra for three of the sources. A faint vertical radio burst was detected in the Stokes V dynamic spectrum of HD 270712. We now go onto discuss the details of the results for each of the four stars in turn.

Brightness temperatures were calculated using equation \ref{eq:bright_temp}, from \cite{dulk1985radio} and also used in \citet{Zic2019}:

\begin{equation}\label{eq:bright_temp}
    T_b = S\left(\frac{c^{2}}{\nu^{2}}\right)\left(\frac{1}{2k}\right)\left(\frac{d^{2}}{A}\right)(1\times10^{-26}),
\end{equation}

\noindent where \textit{S} is the peak flux density in Janskys, $\nu$ is the observed frequency (in this paper the central frequency of 888\;MHz is used), \textit{k} is the Boltzmann constant, \textit{d} is the distance in meters, and \textit{A} is the area of the region where the emission is thought to originate. For the purposes of estimating the brightness temperature the full disk of each star is used as the emitting region. 

\subsection{CD-56 1032A (TIC 220433363)}

Source CD-56 1032A was detected in the second observation field SB25077 data from the 21$^{st}$ March 2021 in Stokes I with 13 hours of integrated images giving a mean flux density of $0.32 \pm 0.01$\;mJy beam$^{-1}$. There was no associated Stokes Q, U nor V emission detected from the source in this observation. The 10-minute integrated light curve in Figure \ref{fig:tess_askap_all} (a) shows a peak radio flux density of 2.89 $\pm$ 0.23\;mJy beam$^{-1}$. 
We calculated a brightness temperature of $2.4\times10^{10} \pm 0.4 \times10^{10} $ K assuming full disk emission using the distance determined from GAIA \citep{Gaia2021}. 

\begin{figure}
	\includegraphics[width=0.9\columnwidth]{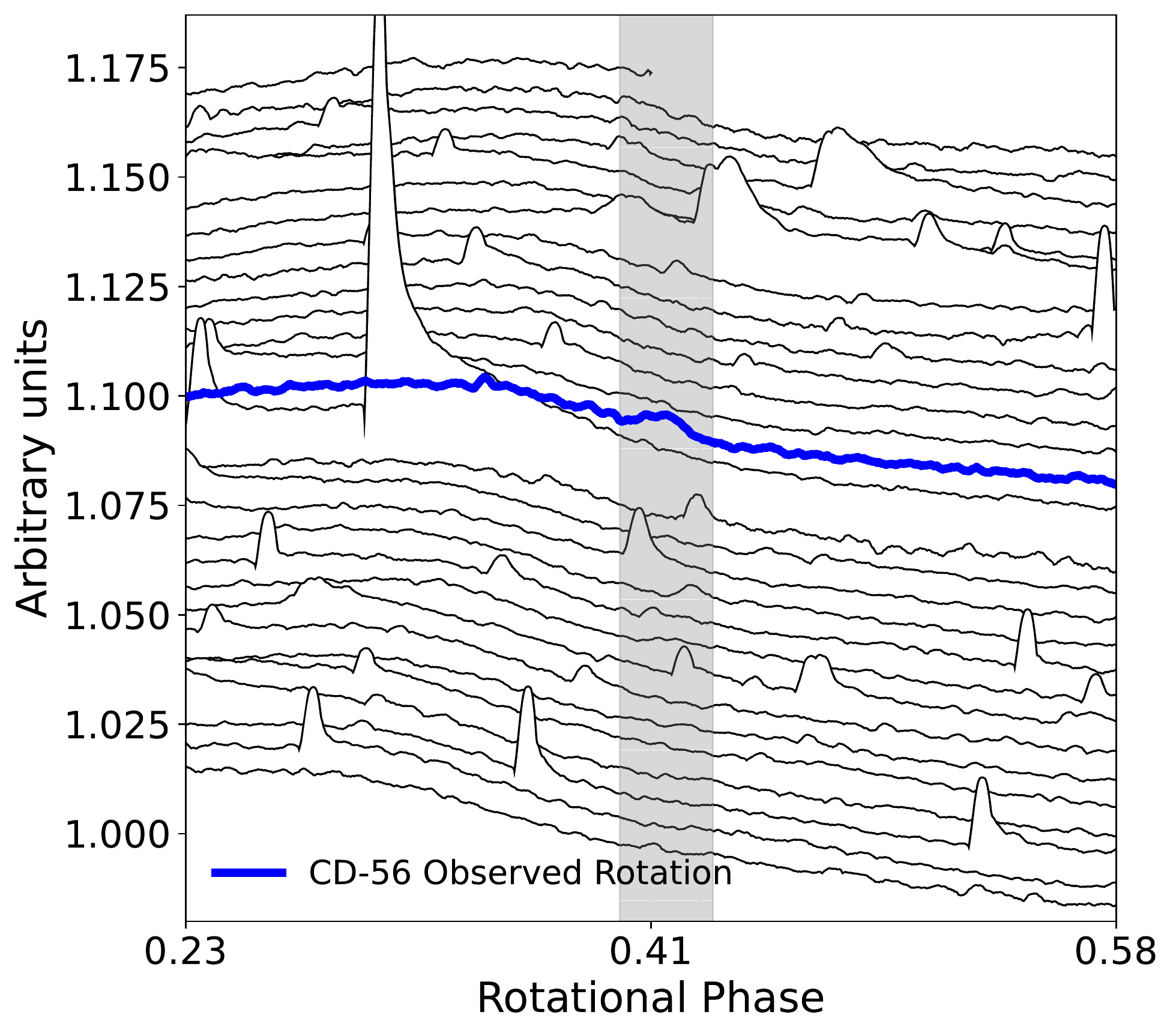}\hfill
    \caption{Folded rotation period of CD-56 1032A (0.8531 days) during the full duration of TESS Sector 36. The blue line corresponds to part of the time co-observed with ASKAP with the raised notch feature highlighted within the overlapping shaded region. The shaded region corresponds to the start and end time of the raised notch. No other feature occurs within this region at the same point of rotation of the source, which may indicate it is not an evolutionary feature of a star spot but rather a single active event.}
    \label{fig:CD56_notch}
\end{figure}

While the radio peak appears to coincide with an increased optical emission in the TESS data, the optical peak was not flagged as a flare event by the automatic flare detection routines. An optical peak at the end of the ASKAP-TESS simultaneous observing period was not detected at radio frequencies. Figure \ref{fig:CD56_notch} was used to examine whether this notch is a persistent star spot feature or an individual transient event.

\subsection{TIC 55497266 (BPS CS 29520-0077)}

TIC 55497266 was also detected in the observation field SB25077. The peak Stokes I flux density from a 13 hour integrated image was $0.25 \pm 0.01$\;mJy\;beam$^{-1}$. Following the same process as carried out for CD-56 1032A, the 10-minute imaging revealed variability with a Stokes I peak flux density of $2.06 \pm 0.14 $\;mJy\;beam$^{-1}$ (see Figure \ref{fig:tess_askap_all}). There was no associated significant Stokes V, Q or U emission detected during the observation. A brightness temperature of $1.4\times10^{11} \pm 0.1 \times10^{11} $\;K was calculated. 
Although the simultaneous ASKAP/TESS observations covered more than one stellar rotation (see Figure \ref{fig:tess_askap_all}), no flares were seen during this time interval.

\subsection{WOH S 2 (TIC 29779873)}

WOH S 2 is one of two active M dwarfs detected in the earlier ASKAP observation (SB25035) on 2021-03-20. A Stokes I 13 hour integrated peak flux density of $0.25\pm 0.03$\;mJy\;beam$^{-1}$ was determined where the peak flux density of the 10-minute integrated light curve was $1.34 \pm 0.07$ \;mJy\;beam$^{-1}$. Stokes V emission was detected from WOH S 2 (see Figure \ref{fig:ASKAP_allstars} and \ref{fig:tess_askap_all}) with a flux density in the 13 hour integrated image of $0.19 \pm 0.03$ \;mJy\;beam$^{-1}$ and a 10-minute peak of $0.44\pm 0.09$ \;mJy\;beam$^{-1}$. The peak circular polarisation fraction of the burst is $32.8\% \pm 7\%$. All Stokes V emission detected from this source is positive, signifying it is left-hand circularly polarised (LCP). The brightness temperature calculated from these observations is $1.60\times10^{12} \pm 0.16\times10^{12} $\;K. As this value likes above the incoherent emission mechanism, it is plausible that either ECM or gyrosynchrotron could be the emission mechanism. However, gyrosynchrotron cannot be disregarded, as the error on the brightness temperature is high, and the degree of circular polarisation is low.

The overlapping TESS light curve was captured as TESS realigned CCDs on the field after down-linking data, and thus was in coarse pointing mode for the duration of the simultaneous observation. The flux variance is of order $\pm 10 \%$. No significant optical flaring activity occurred during this time.

\subsection{HD 270712 (TIC 294750180)}
\begin{figure}
	\includegraphics[width=\columnwidth]{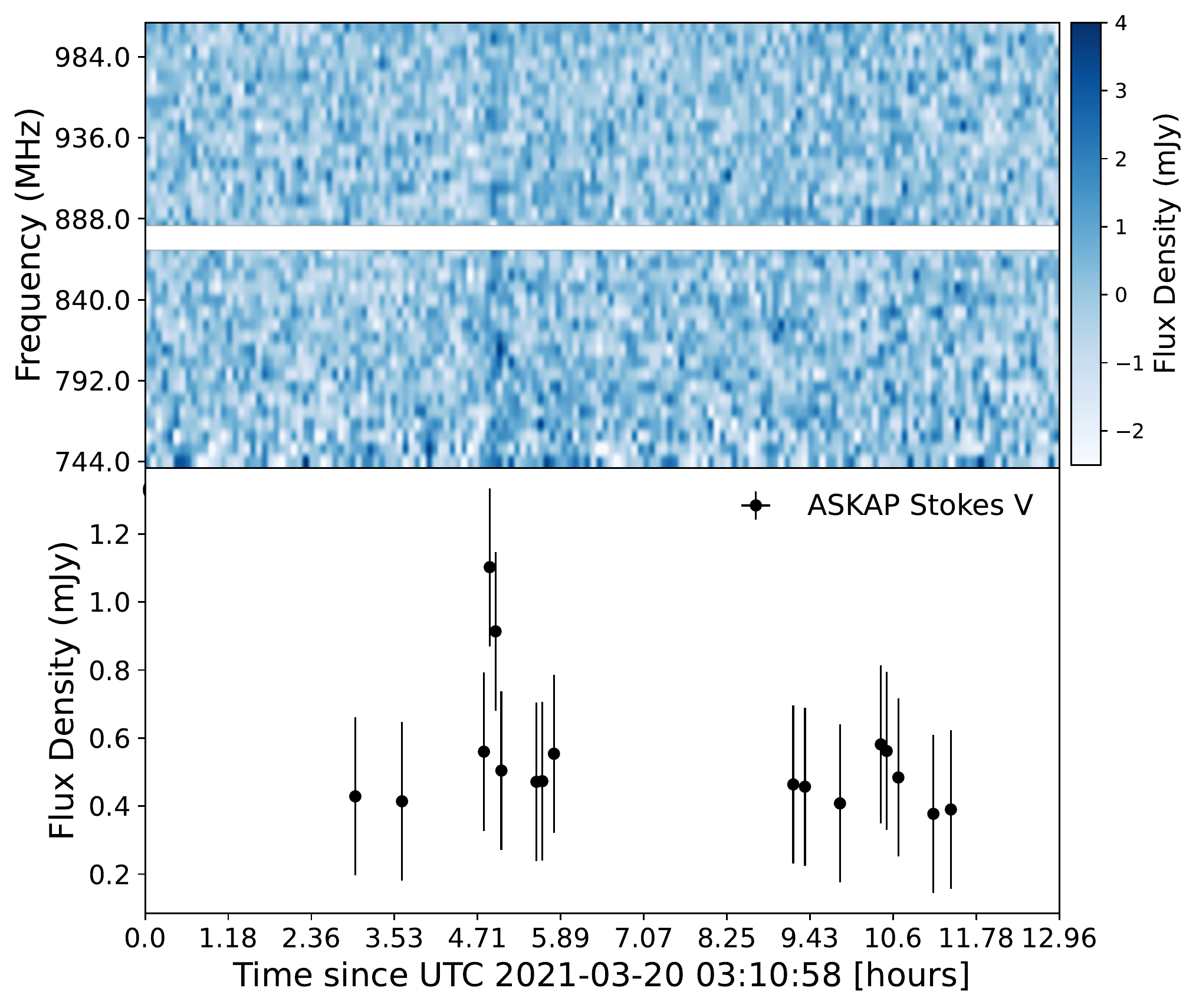}\hfill
    \caption{(Top) ASKAP 5-minute 8\;MHz dynamic spectrum for HD 270712 Stokes V. (Bottom) The $2\sigma$ Stokes V frequency averaged 5-minute time integrated light curve for HD 270712. Note the radio burst at ~4.7 hours after the beginning of simultaneous ASKAP-TESS observations. The integrated flux of the Stokes V burst was measured at $3.2 \sigma$ significance.}
    \label{fig:HD27_dynspec_lightcurve}
\end{figure}

This source was detected in both Stokes I and V during ASKAP observations. HD 270712 is a binary system composing of two active M dwarfs orbiting each other at a separation of \textasciitilde 50\:AU (M0 and M1), both of which may be active \cite{bergfors2010lucky}. The two sources could not be distinguished in either the ASKAP or TESS data. A flare was detected in the ASKAP radio light curve, and was visible across the Stokes V dynamic spectrum as a short duration vertical burst (see figure \ref{fig:HD27_dynspec_lightcurve}). The TESS light curve for this star was corrupted by instrumental effects and was not analysed further. 

There was a weak Stokes I detection for this source in the 13 hour integrated image. A 100\% LCP radio burst was detected. The peak flux density of the 5 minute Stokes V radio light curve was $1.10 \pm 0.01$\;mJy\;beam$^{-1}$. The 13-hour Stokes V integrated flux was $0.15 \pm 0.01$\;mJy\;beam$^{-1}$. The Stokes I integrated flux is $0.20\pm 0.02$\;mJy\;beam$^{-1}$. The radio burst visible in the dynamic spectrum in figure \ref{fig:HD27_dynspec_lightcurve} was not detected in Stokes I. As there is no stellar radius calculation for this source no brightness temperature could be calculated. The peak frequency integrated flux of the Stokes V burst was measured at $3.2 \sigma$ significance, while the peak pixel flux within the burst at 800 MHz from the dynamic spectrum was $\sim2.5 \sigma$ ($4.66\pm 0.02$\;mJy\;beam$^{-1}$).

\subsection{Detection Density}
We detected 4 M dwarfs in the ASKAP data within a two regions of the sky in 13-hour integrated images totalling 60 deg$^{2}$. This is a detection rate of $6.67\times10^{-2}\;$deg$^{-2}$. We compare the detection rate of this work with that reported in \citet{pritchard2021} for K and M dwarfs of $5.27\substack{+3.06 \\ -2.15}\times10^{-4}\;$deg$^{-2}$ at an RMS of 250 $\mu$Jy. Scaling this rate assuming a Euclidean source count distribution, the expected surface density at an RMS of $21\;\mu$Jy is $2.16\times10^{-2}\;$deg$^{-2}$. Comparing this surface density to our rate requires further correction to account for the difference in integration time between our 13-hour observations and the 15-minute Rapid ASKAP Continuum Survey (RACS) Low observations. In addition to the decreased noise threshold and improved sensitivity to faint sources, the longer duration of our observations increases the likelihood of observing short duration bursts. Furthermore, the regions in the current research were chosen for a high number of high cadence TESS active M dwarf targets. We also note that the RACS survey only searched for circularly polarised sources. Two sources in the current research emit with significant Stokes $V$ signals, this gives a Stokes $V$ surface density for the current research of $3.34\times10^{-2}\;$deg$^{-2}$, closer to that of the value determined in \citet{pritchard2021} but still significantly higher, likely due to the reasons mentioned above.

We also compare the current research to the detection levels of M dwarfs in \citet{Callingham2021pop}. Coherent radio emission was detected from 19 M dwarfs in $\sim20$\% of the northern hemisphere (~4125 square degrees). This is a detection density of $4.6\times10^{-3}$ M dwarfs per square degree at 144 MHz. The median sensitivity threshold for these detections was $80\;\mu$Jy. The higher detection density value in the current work may be attributed to the field selection, or the central observed frequency. Furthermore, gyrosynchrotron is more often observed at frequencies $\sim1$\;GHz, while detections at $\sim150$\;MHz (e.g. LOFAR frequencies) are more likely to be attributed to coherent emission processes such as ECM.


\section{Discussion}

The theoretical limit for brightness temperatures from electron cyclotron maser emission (ECM) is $~10^{12}$\;K \cite{dulk1985radio}. When the emission has a brightness temperature higher than this, combined with a high fraction of circular polarisation, it can be classified as ECM \citep{kellermann1969spectra,Melrose1982}. This classification limit has been used to define the emission mechanism powering radio bursts detected from CR Draconis \citep{Callingham2021_crdra}, UV Ceti \citep{Zic2019}, Proxima Centauri \citep{Zic2020}, and many other M dwarfs \citep{Villadsen2019,Callingham2021pop}. The brightness temperatures calculated for three of the detected stars in this study span an order of magnitude surrounding from ; $0.56 \times 10^{11}$\;K (CD-56 1032A), $1.4 \times 10^{11}$\;K (TIC 55497266), and $1.6\times 10^{12}$ K (WOH S 2). These $T_{b}$ values lie both below and just above the limit for coherent emission, thus we cannot conclusively state that the detected radio emission from these sources was produced via the ECM process.

While high circular polarisation is often considered a distinguishing feature of ECM emission, \cite{Hallinan2008} noted that circularly polarised radio emission may not always be detected where ECM is the dominant production mechanism. Radio emission could be depolarised as it passes through the M dwarf magnetic field, as seen in some solar radio bursts \citep{melrose2006depolarization}. Depolarisation also occurs near the solar limb. Reduced polarisation of emission can also be caused by a number of factors highlighted by \cite{Callingham2021_crdra}, which include coronal scattering and reflections passing through density boundaries.

Brightness temperatures below $\sim\,10^{12}$\;K may be indicative of gyrosynchrotron emission \citep{dulk1982simplified}. This incoherent process powered by mildly-relativistic electrons can be detected at the observed frequencies of 1 GHz. A low degree of circular or linear polarisation is often detected from this process. The low degree of circular polarisation detected from WOH S 2 is also supportive of gyrosynchrotron as the dominant process \citep{dulk1985radio, osten2006wide}.

In the case of WOH S 2, a low fraction of circularly polarised radio emission was observed during the radio burst, with a higher brightness temperature ($1.6\times10^{12} \pm 6.7\times10^{11} $\;K) compared to CD-56 1032A and HD 270712. WOH S 2 is a late-type M dwarf with a rotation period 2.5 \textemdash 8 times longer than the other two sources. This may contribute to a lower global magnetic field strength and a reduced ability to depolarise escaping emission. The high brightness temperature and circularly polarised emission detected from WOH S 2 suggest ECM may be the main mechanism. It is possible that a faint optical flare occurred on or near the time of radio emission, however due to the high noise in the TESS light curve it is not possible to verify this. As the brightness temperature lies on the threshold for ECM, within error constraints, and the polarisation fraction is low (<50\%) we cannot rule out an incoherent mechanism such as gyrosynchrotron.

Previous research has highlighted that ECM emission may originate from stellar auroral magnetic activity \citep{Kuz2012, Callingham2021_crdra}. This may be the case with CD-56 1032A and TIC 5549266 as no clear optical counterpart was detected in TESS observations. More radio observations at < 1\;GHz are required to conclude whether ECM or gyrosynchrotron is the primary mechanism powering radio emission from these stars.

TIC 5549266 exhibits peak radio emission aligning with the full starspot on-disk. Bright peak emission was not observed on the first rotation however, so this reduces the likelihood that the starspot is a stable radio emission source. 

The CD-56 1032A TESS optical light curve contains a small yet significant increase in brightness during the radio observation. This emission was discounted as a flare event by {\tt Altaipony}, however it may be evidence for increased activity on the surface of the star which also produced a radio burst. 

The emission from three stars was too faint to produce dynamic spectra, thus there was no detection of signatures suggestive of a drifting radio burst. HD 270712 was detected with a high enough signal-to-noise ratio in Stokes V to produce a dynamic spectrum with a faint vertical burst across the full observed frequency band (see Figure \ref{fig:HD27_dynspec_lightcurve}).

It is possible that the radio waves originate from regions smaller than the full stellar disk used to calculate these values, leading to increased brightness temperature calculations and further supporting ECM as the driving mechanism behind all detected emission. However, without significant circular polarisation detections we cannot conclude that ECM was observed for CD-56 1032A or TIC 55497266. 

Assuming that the brightness temperature calculated is the peak at the time of emission for each source, alternative drivers for CD-56 1032A and TIC 55497266 include plasma and gyrosynchrotron emission. The signature of gyrosynchrotron includes a high brightness temperature with low circular polarisation \citep{dulk1982simplified}, while plasma emission is frequently cited as the main driver of radio emission from solar activity \citep{carley2021observations}. With no supporting clear optical flare observations it is difficult to conclude the emission mechanism of any of the stars from this research.

\section{Summary}

Using simultaneous observations of two regions of the Southern Hemisphere with ASKAP and TESS, we detected faint variable radio emission from four M dwarfs. TESS light curves of the sources did not highlight any significant optical flare events during the observing period. Radio emission characteristics from CD-56 1023A and TIC 55497266 did not match trends in previous research, with no highly circularly polarised radio bursts indicative of flares or possible stellar CMEs detected. The lack of polarisation from these two sources may be due to depolarisation occurring in the strong magnetic fields surrounding the stars. The third source, WOH S 2, contained a relatively high degree of circularly polarised emission which supports previous conclusions that a high energy ECM event powered this emission. A faint circularly polarised radio burst was detected from HD 270712. No optical flare was detected with TESS for any of the sources. This may suggest that a non-flare event accelerated electrons in the stellar corona, perhaps stellar radio noise storms. It is possible that the low signal detected is continuous quiescent emission from these sources with natural variability.

It is likely that the emission originates from regions smaller than the full stellar disk used to calculate brightness temperature values, which would suggest that electron-cyclotron maser is the driving mechanism behind all observed signals. However, without significant circular polarisation detections we cannot conclude that ECM was observed for CD-56 1023A or TIC 55497266. It is also possible that gyrosynchrotron is the mechanism driving the radio emission. Without more information related to the emission (circular or linear polarisation, frequency drift) it is difficult to compare our results to previous M dwarf radio detections, or to determine if any emission is linked to stellar flares or CMEs. Further follow up radio observations with simultaneous high cadence optical data are required to constrain the physical mechanisms driving radio emission from low mass stars.

\section*{Acknowledgements}

The Australian Square Kilometre Array Pathfinder is part of the Australia Telescope National Facility which is managed by CSIRO. 
Operation of ASKAP is funded by the Australian Government with support from the National Collaborative Research Infrastructure Strategy. ASKAP uses the resources of the Pawsey Supercomputing Centre. Establishment of ASKAP, the Murchison Radio-astronomy Observatory, and the Pawsey Supercomputing Centre are initiatives of the Australian Government, with support from the Government of Western Australia and the Science and Industry Endowment Fund. We acknowledge the Wajarri Yamatji as the traditional owners of the Murchison Radio-astronomy Observatory site. Armagh
Observatory and Planetarium is core funded by the Northern Ireland Executive through the Dept. for Communities. This paper includes data collected by the {\tess} mission. Funding for the {\tess} mission is provided by the NASA Explorer Program. The Gaia archive website is \url{https://archives.esac.esa.int/gaia}.
JGD would like to thank the Leverhulme Trust for a Emeritus Fellowship.
JP is supported by Australian Government Research Training Program Scholarship. JR is supported via the Lindsay Fellowship from AOP and DIAS. DK is supported by NSF grant AST-1816492. We would like to thank the referee for their helpful comments and suggestions.

\section*{Data Availability}

The ASKAP data (SB25035 and SB25077) used in this paper can be accessed through the CSIRO ASKAP Science Data Archive (CASDA).
The {\tess} data for each source in this research are available from the NASA MAST portal.



\bibliographystyle{mnras}
\bibliography{askap-tess} 

\begin{thebibliography}{}
\makeatletter
\relax
\def\mn@urlcharsother{\let\do\@makeother \do\$\do\&\do\#\do\^\do\_\do\%\do\~}
\def\mn@doi{\begingroup\mn@urlcharsother \@ifnextchar [ {\mn@doi@}
  {\mn@doi@[]}}
\def\mn@doi@[#1]#2{\def\@tempa{#1}\ifx\@tempa\@empty \href
  {http://dx.doi.org/#2} {doi:#2}\else \href {http://dx.doi.org/#2} {#1}\fi
  \endgroup}
\def\mn@eprint#1#2{\mn@eprint@#1:#2::\@nil}
\def\mn@eprint@arXiv#1{\href {http://arxiv.org/abs/#1} {{\tt arXiv:#1}}}
\def\mn@eprint@dblp#1{\href {http://dblp.uni-trier.de/rec/bibtex/#1.xml}
  {dblp:#1}}
\def\mn@eprint@#1:#2:#3:#4\@nil{\def\@tempa {#1}\def\@tempb {#2}\def\@tempc
  {#3}\ifx \@tempc \@empty \let \@tempc \@tempb \let \@tempb \@tempa \fi \ifx
  \@tempb \@empty \def\@tempb {arXiv}\fi \@ifundefined
  {mn@eprint@\@tempb}{\@tempb:\@tempc}{\expandafter \expandafter \csname
  mn@eprint@\@tempb\endcsname \expandafter{\@tempc}}}

\bibitem[\protect\citeauthoryear{Alvarado-G{\'o}mez, Drake, Cohen, Moschou  \&
  Garraffo}{Alvarado-G{\'o}mez et~al.}{2018}]{alvarado2018}
Alvarado-G{\'o}mez J.~D.,  Drake J.~J.,  Cohen O.,  Moschou S.~P.,   Garraffo
  C.,  2018, The Astrophysical Journal, 862, 93

\bibitem[\protect\citeauthoryear{Alvarado-G{\'o}mez, Drake, Moschou, Garraffo,
  Cohen, Yadav, Fraschetti  et~al.}{Alvarado-G{\'o}mez
  et~al.}{2019}]{alvarado2019coronal}
Alvarado-G{\'o}mez J.~D.,  Drake J.~J.,  Moschou S.~P.,  Garraffo C.,  Cohen
  O.,  Yadav R.~K.,  Fraschetti F.,   et~al., 2019, The Astrophysical Journal
  Letters, 884, L13

\bibitem[\protect\citeauthoryear{Alvarado-G{\'o}mez et~al.,}{Alvarado-G{\'o}mez
  et~al.}{2020}]{alvarado2020tuning}
Alvarado-G{\'o}mez J.~D.,  et~al., 2020, The Astrophysical Journal, 895, 47

\bibitem[\protect\citeauthoryear{{Baraffe}, {Homeier}, {Allard}  \&
  {Chabrier}}{{Baraffe} et~al.}{2015}]{Baraffe2015}
{Baraffe} I.,  {Homeier} D.,  {Allard} F.,   {Chabrier} G.,  2015, \mn@doi
  [\aap] {10.1051/0004-6361/201425481}, \href
  {https://ui.adsabs.harvard.edu/abs/2015A&A...577A..42B} {577, A42}

\bibitem[\protect\citeauthoryear{Bergfors et~al.,}{Bergfors
  et~al.}{2010}]{bergfors2010lucky}
Bergfors C.,  et~al., 2010, Astronomy \& Astrophysics, 520, A54

\bibitem[\protect\citeauthoryear{Bochanski, Hawley, Covey, West, Reid,
  Golimowski  \& Ivezi{\'c}}{Bochanski et~al.}{2010}]{bochanski2010luminosity}
Bochanski J.~J.,  Hawley S.~L.,  Covey K.~R.,  West A.~A.,  Reid I.~N.,
  Golimowski D.~A.,   Ivezi{\'c} {\v{Z}}.,  2010, The Astronomical Journal,
  139, 2679

\bibitem[\protect\citeauthoryear{Callingham et~al.,}{Callingham
  et~al.}{2021a}]{Callingham2021pop}
Callingham J.,  et~al., 2021a, Nature Astronomy, 5, 1233

\bibitem[\protect\citeauthoryear{Callingham et~al.,}{Callingham
  et~al.}{2021b}]{Callingham2021_crdra}
Callingham J.,  et~al., 2021b, Astronomy \& Astrophysics, 648, A13

\bibitem[\protect\citeauthoryear{Carley, Vilmer, Sim{\~o}es  \&
  Fearraigh}{Carley et~al.}{2017}]{carley2017estimation}
Carley E.~P.,  Vilmer N.,  Sim{\~o}es P.~J.,   Fearraigh B.~{\'O}.,  2017,
  Astronomy \& Astrophysics, 608, A137

\bibitem[\protect\citeauthoryear{Carley, Hayes, Murray, Morosan, Shelley,
  Vilmer  \& Gallagher}{Carley et~al.}{2019}]{carley2019loss}
Carley E.~P.,  Hayes L.~A.,  Murray S.~A.,  Morosan D.~E.,  Shelley W.,  Vilmer
  N.,   Gallagher P.~T.,  2019, Nature communications, 10, 1

\bibitem[\protect\citeauthoryear{Carley et~al.,}{Carley
  et~al.}{2021}]{carley2021observations}
Carley E.~P.,  et~al., 2021, The Astrophysical Journal, 921, 3

\bibitem[\protect\citeauthoryear{Cerruti, Kintner~Jr, Gary, Mannucci, Meyer,
  Doherty  \& Coster}{Cerruti et~al.}{2008}]{cerruti2008effect}
Cerruti A.~P.,  Kintner~Jr P.~M.,  Gary D.~E.,  Mannucci A.~J.,  Meyer R.~F.,
  Doherty P.,   Coster A.~J.,  2008, Space Weather, 6

\bibitem[\protect\citeauthoryear{Cornwell, Voronkov  \& Humphreys}{Cornwell
  et~al.}{2012}]{cornwell2012wide}
Cornwell T.,  Voronkov M.,   Humphreys B.,  2012, in Image Reconstruction from
  Incomplete Data VII. p. 85000L

\bibitem[\protect\citeauthoryear{Crosley \& Osten}{Crosley \&
  Osten}{2018}]{Crosley2018}
Crosley M.~K.,  Osten R.~A.,  2018, \mn@doi [The Astrophysical Journal]
  {10.3847/1538-4357/aaaec2}, 856, 39

\bibitem[\protect\citeauthoryear{Crosley et~al.,}{Crosley
  et~al.}{2016}]{crosley2016search}
Crosley M.~K.,  et~al., 2016, The Astrophysical Journal, 830, 24

\bibitem[\protect\citeauthoryear{{Davenport}}{{Davenport}}{2016}]{JDavenport2016}
{Davenport} J. R.~A.,  2016, \mn@doi [\apj] {10.3847/0004-637X/829/1/23}, \href
  {https://ui.adsabs.harvard.edu/abs/2016ApJ...829...23D} {829, 23}

\bibitem[\protect\citeauthoryear{Dulk}{Dulk}{1985}]{dulk1985radio}
Dulk G.~A.,  1985, Annual review of astronomy and astrophysics, 23, 169

\bibitem[\protect\citeauthoryear{Dulk \& Marsh}{Dulk \&
  Marsh}{1982}]{dulk1982simplified}
Dulk G.,  Marsh K.,  1982, The Astrophysical Journal, 259, 350

\bibitem[\protect\citeauthoryear{Fionnag{\'a}in et~al.,}{Fionnag{\'a}in
  et~al.}{2022}]{fionnagain2022coronal}
Fionnag{\'a}in D.~{\'O}.,  et~al., 2022, The Astrophysical Journal, 924, 115

\bibitem[\protect\citeauthoryear{Gaia~Collaboration}{Gaia~Collaboration}{2018}]{gaia2018}
Gaia~Collaboration Brown A. G. A. e.~a.,  2018, \aap, 616, A1

\bibitem[\protect\citeauthoryear{{Gaia Collaboration} et~al.,}{{Gaia
  Collaboration} et~al.}{2021}]{Gaia2021}
{Gaia Collaboration} et~al., 2021, \mn@doi [\aap]
  {10.1051/0004-6361/202039657}, \href
  {https://ui.adsabs.harvard.edu/abs/2021A&A...649A...1G} {649, A1}

\bibitem[\protect\citeauthoryear{G{\"{u}}nther et~al.,}{G{\"{u}}nther
  et~al.}{2020}]{Gunther2020}
G{\"{u}}nther M.~N.,  et~al., 2020, \mn@doi [The Astronomical Journal]
  {10.3847/1538-3881/ab5d3a}, 159, 60

\bibitem[\protect\citeauthoryear{Hallinan, Antonova, Doyle, Bourke, Lane  \&
  Golden}{Hallinan et~al.}{2008}]{Hallinan2008}
Hallinan G.,  Antonova A.,  Doyle J.~G.,  Bourke S.,  Lane C.,   Golden A.,
  2008, \mn@doi [The Astrophysical Journal] {10.1086/590360}, 684, 644

\bibitem[\protect\citeauthoryear{Henry, Jao, Subasavage, Beaulieu, Ianna, Costa
   \& M{\'e}ndez}{Henry et~al.}{2006}]{henry2006solarneighbourhood}
Henry T.~J.,  Jao W.-C.,  Subasavage J.~P.,  Beaulieu T.~D.,  Ianna P.~A.,
  Costa E.,   M{\'e}ndez R.~A.,  2006, The Astronomical Journal, 132, 2360

\bibitem[\protect\citeauthoryear{Hotan et~al.,}{Hotan
  et~al.}{2021}]{hotan2021australian}
Hotan A.,  et~al., 2021, Publications of the Astronomical Society of Australia,
  38

\bibitem[\protect\citeauthoryear{Johnston et~al.,}{Johnston
  et~al.}{2008}]{johnston2008}
Johnston S.,  et~al., 2008, Experimental astronomy, 22, 151

\bibitem[\protect\citeauthoryear{Kellermann \& Pauliny-Toth}{Kellermann \&
  Pauliny-Toth}{1969}]{kellermann1969spectra}
Kellermann K.,  Pauliny-Toth I.,  1969, The Astrophysical Journal, 155, L71

\bibitem[\protect\citeauthoryear{Kochukhov}{Kochukhov}{2021}]{kochukhov2021}
Kochukhov O.,  2021, The Astronomy and Astrophysics Review, 29, 1

\bibitem[\protect\citeauthoryear{{Kuznetsov}, {Doyle}, {Yu}, {Hallinan},
  {Antonova}  \& {Golden}}{{Kuznetsov} et~al.}{2012}]{Kuz2012}
{Kuznetsov} A.~A.,  {Doyle} J.~G.,  {Yu} S.,  {Hallinan} G.,  {Antonova} A.,
  {Golden} A.,  2012, \mn@doi [\apj] {10.1088/0004-637X/746/1/99}, \href
  {https://ui.adsabs.harvard.edu/abs/2012ApJ...746...99K} {746, 99}

\bibitem[\protect\citeauthoryear{{Lightkurve Collaboration}
  et~al.,}{{Lightkurve Collaboration} et~al.}{2018}]{lightkurve2018}
{Lightkurve Collaboration} et~al., 2018, {Lightkurve: Kepler and TESS time
  series analysis in Python} (\mn@eprint {ascl} {1812.013})

\bibitem[\protect\citeauthoryear{Liu et~al.,}{Liu et~al.}{2018}]{liu2018solar}
Liu H.,  et~al., 2018, Solar Physics, 293, 1

\bibitem[\protect\citeauthoryear{Lynch, Lenc, Kaplan, Murphy  \&
  Anderson}{Lynch et~al.}{2017}]{Lynch2017}
Lynch C.~R.,  Lenc E.,  Kaplan D.~L.,  Murphy T.,   Anderson G.~E.,  2017,
  \mn@doi [The Astrophysical Journal] {10.3847/2041-8213/aa5ffd}, 836, L30

\bibitem[\protect\citeauthoryear{Maguire, Carley, McCauley  \&
  Gallagher}{Maguire et~al.}{2020}]{maguire2020}
Maguire C.~A.,  Carley E.~P.,  McCauley J.,   Gallagher P.~T.,  2020, Astronomy
  \& Astrophysics, 633, A56

\bibitem[\protect\citeauthoryear{Marques, Zarka, Echer, Ryabov, Alves, Denis
  \& Coffre}{Marques et~al.}{2017}]{marques2017}
Marques M.,  Zarka P.,  Echer E.,  Ryabov V.,  Alves M.,  Denis L.,   Coffre
  A.,  2017, Astronomy \& Astrophysics, 604, A17

\bibitem[\protect\citeauthoryear{McConnell et~al.,}{McConnell
  et~al.}{2020}]{mcconnell2020RACS}
McConnell D.,  et~al., 2020, Publications of the Astronomical Society of
  Australia, 37

\bibitem[\protect\citeauthoryear{McMullin, Waters, Schiebel, Young  \&
  Golap}{McMullin et~al.}{2007}]{mcmullin2007casa}
McMullin J.~P.,  Waters B.,  Schiebel D.,  Young W.,   Golap K.,  2007, in
  Astronomical data analysis software and systems XVI. p.~127

\bibitem[\protect\citeauthoryear{Melrose}{Melrose}{2006}]{melrose2006depolarization}
Melrose D.,  2006, \apj, 637, 1113

\bibitem[\protect\citeauthoryear{{Melrose} \& {Dulk}}{{Melrose} \&
  {Dulk}}{1982}]{Melrose1982}
{Melrose} D.,  {Dulk} G.,  1982, \apj, 259, 844

\bibitem[\protect\citeauthoryear{{Metodieva}, {Kuznetsov}, {Antonova}, {Doyle},
  {Ramsay}  \& {Wu}}{{Metodieva} et~al.}{2017}]{Met2017}
{Metodieva} Y.~T.,  {Kuznetsov} A.~A.,  {Antonova} A.~E.,  {Doyle} J.~G.,
  {Ramsay} G.,   {Wu} K.,  2017, \mn@doi [\mnras] {10.1093/mnras/stw2597},
  \href {https://ui.adsabs.harvard.edu/abs/2017MNRAS.465.1995M} {465, 1995}

\bibitem[\protect\citeauthoryear{Morosan, Kilpua, Carley  \& Monstein}{Morosan
  et~al.}{2019}]{morosan2019variable}
Morosan D.~E.,  Kilpua E.~K.,  Carley E.~P.,   Monstein C.,  2019, Astronomy \&
  Astrophysics, 623, A63

\bibitem[\protect\citeauthoryear{Morosan, Kumari, Kilpua  \& Hamini}{Morosan
  et~al.}{2021}]{morosan2021moving}
Morosan D.,  Kumari A.,  Kilpua E.,   Hamini A.,  2021, Astronomy \&
  Astrophysics, 647, L12

\bibitem[\protect\citeauthoryear{Moschou, Drake, Cohen, Alvarado-G{\'o}mez,
  Garraffo  \& Fraschetti}{Moschou et~al.}{2019}]{moschou2019stellar}
Moschou S.-P.,  Drake J.~J.,  Cohen O.,  Alvarado-G{\'o}mez J.~D.,  Garraffo
  C.,   Fraschetti F.,  2019, The Astrophysical Journal, 877, 105

\bibitem[\protect\citeauthoryear{Mullan \& Paudel}{Mullan \&
  Paudel}{2019}]{mullan2019}
Mullan D.,  Paudel R.,  2019, The Astrophysical Journal, 873, 1

\bibitem[\protect\citeauthoryear{Murphy et~al.,}{Murphy
  et~al.}{2013}]{murphy2013vast}
Murphy T.,  et~al., 2013, Publications of the Astronomical Society of
  Australia, 30

\bibitem[\protect\citeauthoryear{{Murphy} et~al.,}{{Murphy}
  et~al.}{2021}]{Murphy2021}
{Murphy} T.,  et~al., 2021, \mn@doi [\pasa] {10.1017/pasa.2021.44}, \href
  {https://ui.adsabs.harvard.edu/abs/2021PASA...38...54M} {38, e054}

\bibitem[\protect\citeauthoryear{Nelson \& Melrose}{Nelson \&
  Melrose}{1985}]{nelson1985type}
Nelson G.,  Melrose D.,  1985, Solar radiophysics: Studies of emission from the
  Sun at metre wavelengths, pp 333--359

\bibitem[\protect\citeauthoryear{Osten \& Bastian}{Osten \&
  Bastian}{2006}]{osten2006wide}
Osten R.~A.,  Bastian T.,  2006, The Astrophysical Journal, 637, 1016

\bibitem[\protect\citeauthoryear{Osten \& Wolk}{Osten \&
  Wolk}{2015}]{osten2015connecting}
Osten R.~A.,  Wolk S.~J.,  2015, The Astrophysical Journal, 809, 79

\bibitem[\protect\citeauthoryear{Osten et~al.,}{Osten et~al.}{2010}]{Osten2010}
Osten R.~A.,  et~al., 2010, \mn@doi [Astrophysical Journal]
  {10.1088/0004-637X/721/1/785}, 721, 785

\bibitem[\protect\citeauthoryear{{Pecaut} \& {Mamajek}}{{Pecaut} \&
  {Mamajek}}{2013}]{PecautMamajek2013}
{Pecaut} M.~J.,  {Mamajek} E.~E.,  2013, \mn@doi [\apjs]
  {10.1088/0067-0049/208/1/9}, \href
  {https://ui.adsabs.harvard.edu/abs/2013ApJS..208....9P} {208, 9}

\bibitem[\protect\citeauthoryear{Pope, Callingham, Feinstein, G{\"u}nther,
  Vedantham, Ansdell  \& Shimwell}{Pope et~al.}{2021}]{pope2021tess}
Pope B.~J.,  Callingham J.~R.,  Feinstein A.~D.,  G{\"u}nther M.~N.,  Vedantham
  H.~K.,  Ansdell M.,   Shimwell T.~W.,  2021, The Astrophysical Journal
  Letters, 919, L10

\bibitem[\protect\citeauthoryear{Pritchard et~al.,}{Pritchard
  et~al.}{2021}]{pritchard2021}
Pritchard J.,  et~al., 2021, Monthly Notices of the Royal Astronomical Society,
  502, 5438

\bibitem[\protect\citeauthoryear{Raja et~al.,}{Raja
  et~al.}{2022}]{raja2022spectral}
Raja K.~S.,  et~al., 2022, The Astrophysical Journal, 924, 58

\bibitem[\protect\citeauthoryear{Rajpurohit, Reyl{\'e}, Allard, Homeier,
  Schultheis, Bessell  \& Robin}{Rajpurohit
  et~al.}{2013}]{rajpurohit2013effective}
Rajpurohit A.,  Reyl{\'e} C.,  Allard F.,  Homeier D.,  Schultheis M.,  Bessell
  M.,   Robin A.,  2013, Astronomy \& Astrophysics, 556, A15

\bibitem[\protect\citeauthoryear{{Ramsay}, {Kolotkov}, {Doyle}  \&
  {Doyle}}{{Ramsay} et~al.}{2021}]{ramsay2021}
{Ramsay} G.,  {Kolotkov} D.,  {Doyle} J.~G.,   {Doyle} L.,  2021, \mn@doi
  [\solphys] {10.1007/s11207-021-01899-x}, \href
  {https://ui.adsabs.harvard.edu/abs/2021SoPh..296..162R} {296, 162}

\bibitem[\protect\citeauthoryear{Reid \& Ratcliffe}{Reid \&
  Ratcliffe}{2014}]{reid2014review}
Reid H. A.~S.,  Ratcliffe H.,  2014, Research in Astronomy and Astrophysics,
  14, 773

\bibitem[\protect\citeauthoryear{{Ricker} et~al.,}{{Ricker}
  et~al.}{2015}]{Ricker2015}
{Ricker} G.~R.,  et~al., 2015, \mn@doi [Journal of Astronomical Telescopes,
  Instruments, and Systems] {10.1117/1.JATIS.1.1.014003}, \href
  {http://adsabs.harvard.edu/abs/2015JATIS...1a4003R} {1, 014003}

\bibitem[\protect\citeauthoryear{Schmidt et~al.,}{Schmidt
  et~al.}{2019}]{Schmidt2019}
Schmidt S.~J.,  et~al., 2019, \mn@doi [The Astrophysical Journal]
  {10.3847/1538-4357/ab148d}, 876, 115

\bibitem[\protect\citeauthoryear{{Schmitt}, {Ioannidis}, {Robrade}, {Czesla}
  \& {Schneider}}{{Schmitt} et~al.}{2019}]{Schmitt2019}
{Schmitt} J.~H.~M.~M.,  {Ioannidis} P.,  {Robrade} J.,  {Czesla} S.,
  {Schneider} P.~C.,  2019, \mn@doi [\aap] {10.1051/0004-6361/201935374}, \href
  {https://ui.adsabs.harvard.edu/abs/2019A&A...628A..79S} {628, A79}

\bibitem[\protect\citeauthoryear{{Stassun} et~al.,}{{Stassun}
  et~al.}{2019}]{Stassun2019}
{Stassun} K.~G.,  et~al., 2019, \mn@doi [\aj] {10.3847/1538-3881/ab3467}, \href
  {https://ui.adsabs.harvard.edu/abs/2019AJ....158..138S} {158, 138}

\bibitem[\protect\citeauthoryear{Stepanov, Kliem, Zaitsev, F{\"u}rst, Jessner,
  Kr{\"u}ger, Hildebrandt  \& Schmitt}{Stepanov
  et~al.}{2001}]{stepanov2001microwave}
Stepanov A.,  Kliem B.,  Zaitsev V.,  F{\"u}rst E.,  Jessner A.,  Kr{\"u}ger
  A.,  Hildebrandt J.,   Schmitt J.,  2001, Astronomy \& Astrophysics, 374,
  1072

\bibitem[\protect\citeauthoryear{Vedantham}{Vedantham}{2021}]{vedantham2021mechanism}
Vedantham H.,  2021, Monthly Notices of the Royal Astronomical Society, 500,
  3898

\bibitem[\protect\citeauthoryear{Vedantham et~al.,}{Vedantham
  et~al.}{2020}]{vedantham2020coherent}
Vedantham H.,  et~al., 2020, Nature Astronomy, 4, 577

\bibitem[\protect\citeauthoryear{Villadsen \& Hallinan}{Villadsen \&
  Hallinan}{2019}]{Villadsen2019}
Villadsen J.,  Hallinan G.,  2019, \mn@doi [The Astrophysical Journal]
  {10.3847/1538-4357/aaf88e}, 871, 214

\bibitem[\protect\citeauthoryear{Wang, Tuntsov, Murphy, Lenc, Walker,
  Bannister, Kaplan  \& Mahony}{Wang et~al.}{2021}]{wang2021askap}
Wang Y.,  Tuntsov A.,  Murphy T.,  Lenc E.,  Walker M.,  Bannister K.,  Kaplan
  D.~L.,   Mahony E.~K.,  2021, Monthly Notices of the Royal Astronomical
  Society, 502, 3294

\bibitem[\protect\citeauthoryear{Whiting \& Humphreys}{Whiting \&
  Humphreys}{2012}]{whiting2012source}
Whiting M.,  Humphreys B.,  2012, Publications of the Astronomical Society of
  Australia, 29, 371

\bibitem[\protect\citeauthoryear{Zic et~al.,}{Zic et~al.}{2019}]{Zic2019}
Zic A.,  et~al., 2019, Monthly Notices of the Royal Astronomical Society, 488,
  559

\bibitem[\protect\citeauthoryear{Zic et~al.,}{Zic et~al.}{2020}]{Zic2020}
Zic A.,  et~al., 2020, The Astrophysical Journal, 905, 23

\makeatother
\end{thebibliography}




\appendix


\bsp	
\label{lastpage}
\end{document}